\documentclass[10pt,twoside,a4paper]{article} 
 
\usepackage{amsmath, amsfonts, amsthm, amssymb, amscd}
\usepackage[mathcal]{euscript} 
\usepackage{graphicx} 
\usepackage{dsfont, pxfonts, mathrsfs, accents, verbatim}
\usepackage{chngpage}
\theoremstyle{plain}

       \numberwithin{equation}{section}

\usepackage{amsfonts} 
\usepackage{a4wide}
\usepackage{epsfig}
\usepackage{graphicx}
\usepackage{subfig}  

\newcommand \tT         {\widetilde T} 
\newcommand \del            \partial 
\newcommand \RR             {\mathbb  R} 

\newcommand \la \langle
\newcommand \ra \rangle

\newcommand{\auth}{\textsc}
\newcommand \be         {\begin{equation}}
\newcommand \ee         {\end{equation}}

\newcommand \TT     {\mathcal{T}}
\newcommand \Tcal   {\TT} 
\newcommand \overtT    		{{{{\overline{\widetilde T}}}}}

\newcommand \gt    	    {\widetilde g}

\newcommand \Hcal           {\mathcal{H}}

\newcommand \MM            {\mathbb{M}}

\newcommand \Fp    		 {\mathbf{F}} 
\newcommand \Sp    		 {\mathbf{S}} 
 
\newcommand \Ur    		{\overline  {\mathbf{U}}} 
\newcommand \Ubar   		{\overline  {U}} 
\newcommand \Fbar   		{\overline  {F}} 
\newcommand \Sbar   		{\overline  {S}}

\newcommand \Mcal 	{\mathcal M}

\newcommand \bei             {\begin{itemize}}
\newcommand \eei             {\end{itemize}}

\usepackage{amsmath, amsfonts, amsthm, amssymb, amscd}
\usepackage{graphicx}  
\usepackage{a4wide}
\usepackage[mathcal]{euscript}  
\usepackage{subfig} 
\numberwithin{equation}{section}
\let\oldmarginpar\marginpar
\renewcommand\marginpar[1]{\-\oldmarginpar[\raggedleft\footnotesize #1]%
{\raggedright\footnotesize #1}}
\newcommand \Kb {\overline K}

\newcommand \nablat {\widetilde \nabla}
\newcommand \hatT {\widehat T} 

\newcommand \Up    		 {\mathbf{U}}  

\newcommand \Sb   		{\overline S}

\newcommand \uN 		{\overline  {N} } 
\newcommand \uT   		{\overline T} 
 
\newcommand \Fr    		{\overline  {\mathbf{F}}} 
\newcommand \Sr    		{\overline  {\mathbf{S}}} 
\newcommand \sgn {\text{sgn}}
\begin{document}

\title{The geometric finite volume method for compressible fluid flows on Schwarzschild spacetime} 
\author{ 
Philippe G. LeFloch$^*$ and Hasan Makhlof\footnote{Laboratoire Jacques--Louis Lions \& Centre National de la Recherche Scientifique,
Universit\'e Pierre et Marie Curie (Paris 6), 4 Place Jussieu, 75252 Paris, France.
Email: {\sl contact@philippelefloch.org, makhlof@ann.jussieu.fr}.
\newline
\textit{\ AMS Subject Class.} 76L05, 35L65. 
\textit{Keywords and Phrases.} Compressible fluid, Euler system, Schwarzschild spacetime, geometric formulation, 
finite volume method, steady state.  
}
}
\date{\today}
\maketitle

 \begin{abstract} 
 We consider the relativistic Euler equations governing spherically symmetric, perfect fluid flows on the outer domain of communication of Schwarzschild spacetime, and we introduce a version of the finite volume method which is formulated geometrically (without choosing coordinates a priori) 
and is well--balanced, in the sense that it preserves steady solutions to the Euler equations on the curved geometry under consideration. In order to formulate our method, we first derive a closed formula describing all steady and spherically symmetric solutions to the Euler equations posed on Schwarzschild spacetime.  Second, we describe a finite volume method which is formulated geometrically from the family of steady solutions to the Euler system.  Our scheme is second--order accurate and, as required, preserves the family of steady solutions at the discrete level. Numerical experiments are presented which demonstrate the efficiency and robustness of the proposed method even for solutions containing shock waves and nonlinear interacting 
wave patterns. As an application, we investigate the late--time asymptotics of perturbed steady solutions and demonstrate its convergence for late time toward another steady solution, taking the overall effect of the perturbation into account. 
 \end{abstract}

\tableofcontents


 \section{Introduction}

The finite volume method is a versatile technique for scientific computing, which has found many applications in physical and engineering sciences. In particular, it allows one to approximate weak solutions (containing shock waves) to nonlinear hyperbolic systems of balance laws such as, for instance, the Euler equations of compressible fluid dynamics. In the present paper, we propose a geometric version of the finite volume method for general balance laws of hyperbolic partial differential equations, and we apply this method to the Euler equations for spherically symmetric, relativistic fluid flows posed on a curved spacetime and, for definiteness, the outer domain of communication of Schwarzschild spacetime. The proposed method is second--order accurate (in smooth regions of the flow), and well--balanced in the sense that steady solutions to the relativistic fluid equations are preserved at the discrete level. 

The balance laws of interest in the present work have the following general form. Given 
a spacetime  $(\Mcal,g)$ with Lorentzian metric $g$ and covariant derivative operator $\nabla$, 
we consider
 the class of balance laws
\be
\label{111}
\nabla_\alpha \big( {T^\alpha}_\beta (\phi) \big) = 0,
\ee 
where ${T^\alpha}_\beta(\phi)$ represents the energy--momentum tensor of an (unknown) tensor field $\phi$ defined on $\Mcal$ (the indices $\alpha, \beta$ ranging between $0$ and $3$). We use a standard notation for the metric 
$g=g_{\alpha\beta} dx^\alpha dx^\beta$ in coordinates $(x^\alpha)$, and repeated indices are implicity summed up. We lower (or raise) indices with the metric $g_{\alpha\beta}$ (or its inverse $g^{\alpha\beta}$) so that, for instance, $u_\alpha = g_{\alpha\beta} u^\beta$ for a vector field $u^\alpha$. 

In particular, we are interested in the relativistic Euler equations for perfect compressible fluids, corresponding to $\phi = (\rho, u)$ 
in \eqref{111} with 
\be
\label{121}
{T^\alpha}_\beta (\rho, u) = (\rho c^2 + p) \, u^\alpha u_\beta + p \, {g^\alpha}_\beta. 
\ee 
Here, the scalar field $\rho \geq 0$ denotes the mass--energy density of the fluid and the vector field $u^\alpha$ its velocity, normalized so that $u^\alpha u_\alpha = -1$, while $c>0$ represents the light speed. Moreover, the pressure function in \eqref{121} is given by an equation of state $p = p(\rho)$ which must satisfy the (hyperbolicity) condition $p'(\rho) \in (0,c)$ (for all $\rho>0$), so that 
the equations \eqref{111} can be written in local coordinates as a system of nonlinear balance laws, which is strictly hyperbolic for $\rho>0$.
In general, initially smooth solutions to \eqref{111}--\eqref{121} become discontinuous in finite time and shock waves form and then 
propagate within the spacetime. 

A broad literature
is available on the design of robust and accurate, shock--capturing schemes for general hyperbolic systems posed on a flat geometry like the Minkowski spacetime. In the present work, we intend to also take a {\sl curved} background geometry into account, by following recent work by the first author and his collaborators; cf.~\cite{ABL,AmorimLeFlochOkutmustur,LeFlochOkutmustur}.  
To this end, we introduce a finite volume scheme which is based on the geometric formulation \eqref{111}, rather than on the corresponding partial differential equations in a specific local coordinate chart. Moreover, in order to achieve the well--balanced property, we extend the approach in Russo et al.  \cite{PuppoRusso,Russo1,Russo2} and LeFloch et al. \cite{LMO},  
and we introduce a discretization which accurately takes into account the family of steady solutions to the balance laws and, therefore, the geometric effects induced by the Lorentzian geometry $(\Mcal,g)$. To implement this strategy, it is necessary to first investigate the class of steady solutions to the Euler system on the curved background under consideration.

Numerical relativity has undergone a tremendous development in recent years, and the reader is referred to Mart\'i and M\"uller \cite{MM} for a 
review of numerical methods developed first for special relativity, and to Font \cite{Font} and Alcubierre \cite{Alcubierre} for a review in the context of general relativity. 
See also Papadopoulos and Font \cite{PapaFont99,PapaFont} and 
Novak and Ib\'a\`nez \cite{NI}.
Various astrophysical applications have been successfully dealt with in recent years, including the evolution of neutron stars and the merging of black holes.  For background on general relativity, we refer the reader to \cite{Wald} and, for
 theoretical and numerical tools concerning
 nonlinear hyperbolic equations and their discretization, we refer to \cite{Dafermos-book,LeFloch-book} and \cite{LeVeque-book}, respectively. 

An outline of this paper is as follows. In Section~\ref{sec:2}, we introduce a version of the finite volume method, which we state fully geometrically on a curved spacetime and for the general balance laws \eqref{111}. In Section~\ref{sec:3}, we consider the relativistic Euler equations when the background is chosen to be the outer domain of communication of Schwarzschild spacetime, and we determine all steady solutions to \eqref{111}--\eqref{121} in this context. Next, in Section~\ref{sec:4}, we introduce a finite volume method for the Euler system, which is well-balanced and second--order accurate. Finally, in Section~\ref{sec:5}, various numerical tests are presented, which demonstrate the efficiency of the proposed scheme;  as an application, we study the late--time asymptotics of steady solutions under perturbation. Section~\ref{section:6} contains concluding remarks and refers the reader to the follow-up paper \cite{LMH} concerned with self-gravitating matter in 
spherical symmetry. 
   

\section [The geometric finite volume method]{The geometric finite volume method on a curved spacetime}
\label{sec:2}

\subsection{Spacetime foliations by spacelike hypersurfaces}    

We begin by presenting a framework based on a general four--dimensional spacetime $(\Mcal,g)$, 
that is, a manifold (possibly with boundary) endowed with a Lorentzian metric $g$ with signature $(-,+,+,+)$. 
We denote by $\nabla$ the spacetime Levi--Civita connection associated with this metric. As is customary, we assume that $\Mcal$ admits a foliation $\big\{ \Hcal_t \big\}_{t \in [0, \infty)}$  by oriented spacelike
 hypersurfaces such that 
the parameter $t: \Mcal \to \RR_+$ provides us with a global time function, satisfying $dt \neq 0$ on $\Mcal$. This allows us  
to distinguish between future-oriented ($t$ increasing) and past--oriented ($t$ decreasing) timelike directions on $\Mcal$. 
By definition, we thus have 
$\Mcal = \bigcup_{t \geq 0} \Hcal_t$, 
and $\Mcal$ is topologically diffeomorphic to $\RR_+\times \Hcal_0$ while its boundary is the union of
the initial slice $\Hcal_0$ and a boundary $\RR_+ \times \del \Hcal_0$ determined from $\del \Hcal_0$, i.e. 
$$
\Mcal \cup \del \Mcal = \bigcup_{t \geq 0} \overline\Hcal_t, 
\qquad
\qquad
 \overline\Hcal_t = \Hcal_t \cup \del \Hcal_t. 
$$ 
(Observe that the spatial slices may have a non--trivial boundary.)  

Given local coordinates $(x^\alpha) = (t, x^j)$, in which Greek indices describe $0, \ldots, 3$ while Latin indices describe $1,2,3$, 
we express the spacetime metric in the form $g = g_{\alpha\beta} \, dx^\alpha dx^\beta$, 
and we denote by $(g^{\alpha\beta})$ its inverse. The assumed $(3+1)$--decomposition of the spacetime is standard in general relativity (cf., for instance, the textbooks~\cite{Alcubierre,Wald}) and is determined by the time function $t$. We also choose coordinates $(x^j)$ on the initial slice $\Hcal_0$ and propagate them in the spacetime along the vector field $\nabla t$. This leads us to the metric decomposition 
\be
\label{ThreeOne}
ds^2 = - N^2 dt^2 + \gt,  
\ee
where $N =(-g( dt, dt))^{-1/2} >0$ is referred to as the lapse function and $\gt =\gt_t = g_{ij} \, dx^i dx^j$ represents the induced Riemannian metric on the slices. Denoting by $dV_g$ and $dV_{\gt_t}$ the volume forms associated with the Lorentzian and Riemannian metrics, respectively, we can write 
\be
\label{ThreeOne-2}
dV_g = N \, dt dV_{\gt}, \qquad   B = \big(\det \gt\big)^{1/2}, \qquad NB = (\det g)^{1/2}.  
\ee

Let $n= N\,\nabla t$ be the future directed, timelike unit normal to the slices and $K$ be the second fundamental form, defined by 
$K (X,Y) = - g(\nabla_X n, Y)$ for all vectors $X,Y$ tangent to the hypersurface $\Hcal_t$, so that
$$
K_{ij} = - {1 \over 2N} \del_0 g_{ij}
$$
in local coordinates $(t, x^j)$ adapted to the foliation. We introduce also the Levi-Civita connection $\nablat$ of the slices $(\Hcal_t, \gt_t)$, given (for any tangent vector fields $X,Y$) by 
$$
\nablat_Y X = \nabla_Y X + K(X,Y) \, n.
$$
Finally, the Christoffel symbols 
$\Gamma_{\alpha\beta}^\gamma ={1\over 2} \, g^{\gamma \theta}\big( 
\del_\alpha g_{\beta\theta} + \del_\beta g_{\alpha\theta} - \del_\theta g_{\alpha\beta}\big)$ of the spacetime metric read
 \be
\label{gam1}
\aligned
&\Gamma_{0\alpha}^0 = {\del_\alpha N \over N}, 
\hskip2.cm 
\Gamma_{00}^j = {1 \over 2} g^{jk} \del_k N^2, 
\\
&\Gamma_{0i}^j = {1 \over 2} g^{jk} \del_0 g_{ik},   
\hskip1.3cm 
\Gamma_{ij}^0 = {1 \over 2 N^2} \del_0 g_{ij}, 
\qquad \quad 
\Gamma_{j\alpha}^j = {\del_\alpha B \over B}. 
\endaligned
\ee


\subsection{Spacetime formulation of balance laws}

Consider now the general system of balance laws \eqref{111} and, by first assuming enough regularity on the solutions, let us rewrite it 
in local coordinates adapted to the $(3+1)$--decomposition \eqref{ThreeOne} determined by the time function $t$. From the definition of the covariant derivative,  \eqref{111}  is equivalent to 
$$
\aligned
\del_{0} T^{0\beta}+ \del_{j} T^{j\beta} 
+ \Gamma^0_{00}T^{0\beta} 
& + \Gamma^j_{j0}T^{\beta0}
+\Gamma^0_{0j}T^{j\beta}+\Gamma^j_{jk}T^{k\beta} 
\\
& + \Gamma^\beta_{00}T^{00}
+2 \, \Gamma^\beta_{j0}T^{j0}+\Gamma^\beta_{jk}T^{jk} = 0  
\endaligned
$$
or, in view of the expressions of the Christoffel symbols \eqref{gam1},   
$$
\aligned
 \del_0 T^{00} + \del_j T^{j0}  
&= - \del_\alpha (\ln (BN^3)) \, T^{0\alpha} - \del_0 (\ln N) \,  T^{00} + {1 \over N} K_{ij} \,  T^{ij}, 
\\
\del_0 T^{0i} + \del_j T^{ij}
& = - \del_\alpha (\ln (BN^3)) \, T^{i\alpha}   
       + {1\over 2} g^{ik} \del_k (N^2) \, T^{00} 
 \\
& \quad      + 2N \, g^{ik} K_{jk} \,  T^{j0} 
       - \Gamma^i_{jk} \,  T^{jk}
      + 2 \, {\del_0 N \over N} \, T^{0i}
       + 2 \, {\del_j B \over B} \,  T^{ij}. 
\endaligned
$$
After multiplication by the weight $BN^3$, we obtain the following formulation of the balance laws  
\be
\label{cou1}
\aligned
\del_0({BN^3}T^{0 0})+\del_j({BN^3}T^{0j})&=S^0,
\\
 \del_0 (BN^3T^{i0})+\del_j (BN^3T^{ij})&=S^i,
\endaligned
\ee
with right-hand sides 
$$
\aligned
S^0=& \, BN^2 \, \del_0 N \, T^{00} - {BN\over 2} \del_0 g_{kj} \, T^{kj},
\\
S^i=& \, {BN^3\over 2} g^{ik} \del_kN^2 \, T^{00} - BN^3 \, g^{ik} \del_0g_{jk} \, T^{j0} - BN^3 \, \Gamma^i_{jk}T^{jk}
\\
&  -T^{0i}\del_0(BN^3)
+2T^{0i}BN^2\del_0 N+2N^3T^{ij}\del_j B.
\endaligned
$$
In particular, plugging in \eqref{cou1} the expression of the matter tensor  \eqref{121} yields the formulation of the Euler equations on a curved spacetime: 
\be
\label{ecur}
\aligned
& \del_0 \Big(BN^3 \Big( (\rho c^2 + p) \, u^0 \, u^0 - {1\over N^2} p \Big) \Big)
   + \del_j \Big(BN^3\, (\rho c^2 + p) \, u^0 \, u^j \Big) 
   = S^0, 
\\
& \del_0 \Big( BN^3 (\rho c^2 + p) \, u^0 \, u^i \Big) 
 + \del_j \Big(BN^3\Big(  (\rho c^2 + p) \, u^i \, u^j  +  p \, g^{ij} \Big) \Big) = S^i, 
\endaligned
\ee 
which consist of four equations for the five unknowns $(\rho, u^\alpha)$, satisfying the  
 constraint $u^\alpha u_\alpha = -1$.

Introducing local coordinates $(x^\alpha)= (t, x^j)$ and recalling the decomposition \eqref{ThreeOne-2} of the volume $dV_g$, the formulation \eqref{ecur} can be recovered in the sense of distributions. 
 

\subsection{The geometric spacetime finite volume method} 
\label{sec:volfini}
 
We are now in a position to introduce the geometric formulation of the finite volume method, which we design directly from the covariant form \eqref{111}
of the balance laws, rather than introducing first coordinates and then a discretization.

First of all, we introduce a triangulation of the spacetime, say 
$$
\Mcal = \bigcup_{K \in \Tcal} \Kb,    
$$
made of finitely many open sets $K$, which are assumed to satisfy the following conditions:
\begin{itemize}

\item The boundary $\del K = \bigcup_{e\subset \del K} e$ is  piecewise smooth and contains two spacelike faces (i.e., having an induced metric of Riemannian type) denoted by  $e_K^+$ and $e_K^-$, and timelike (or ``vertical'') faces (i.e.~having an induced metric of Lorentzian type), the latter faces being denoted by 
$$
e^0 \in \del^0 K = \del K \setminus \big\{e_K^+, e_K^-\big\}.
$$

\item The intersection $\Kb \cap \Kb'$ of two distinct elements is  a common face of $K$ and $K'$, or else is a smooth submanifold with dimension at most $2$.
\end{itemize}

 \noindent We then adopt the following notation: 

\begin{itemize}

\item Along the timelike faces $e_K^\pm$, we introduce the outgoing unit normal vector field denoted by $n_K$. 

\item $|K|,|e_K^+|,|e_K^-|, |e^0|$ denote the Lebesgue measure of the sets $K,e_K^+, e_K^-,e^0$, respectively, 
which is defined from the Lorentzian metric or the induced metric on these hypersurfaces.

\end{itemize}

Furthermore, when the spacetime is endowed with a foliation by spacelike hypersurfaces, say
$\Mcal \cup \del \Mcal = \bigcup_{t\geq0} \overline\Hcal_t$, associated with a time function $t$ : $\Mcal \bigcup\del \Mcal \to [0, +\infty)$
we say that the triangulation 
$$
\Mcal \cup \del\Mcal = \bigcup_{K \in \Tcal} \Kb,    
$$
is {\sl compatible with the time function} $t$ if it is determined from a sequence of discrete times
$$
t_0, \, t_1, \, t_2, \ldots
$$
and a triangulation $\Tcal'$ of the initial three--dimensional slice $\Hcal_0$, say
$$
\Hcal_0  = \bigcup_{K \in \Tcal'} \Kb',    
$$
in such a way that the boundaries of the elements $\Kb'$ are 
transported to the whole spacetime along the vector field $\del_t$ associated with 
the time function so that all the vertical faces are parallel to this vector field. 
This property makes it clearer to advance the numerical solution forward in time, and is assumed from now on. 

The finite volume method is then based on the following general weak form of the system of balance laws. Recall that solutions to
(nonlinear hyperbolic) balance laws are generally discontinuous, and these equations must be understood in the 
sense of distributionals. 
Hence, we seek here for {\sl weak solutions} for which \eqref{111} is understood in the averaged sense 
\be
\label{Eulerfaible}
\int_\Mcal \pi^{(X)}_{\alpha\beta} \, T^{\alpha\beta} \, dV_g = 0,
\ee
in which $X^\alpha$ is a test-field (i.e.~is smooth and compactly supported in $\Mcal$)
and $\pi^{(X)}$ denotes its deformation tensor defined by
\be
\label{eq:pi}
\pi^{(X)}_{\alpha \beta} 
= {1 \over 2} \big( \nabla_\alpha X_\beta + \nabla_\beta X_\alpha \big).
\ee

The finite volume method is based on the above integral formula, except that we must now integrate over an arbitrary spacetime element $K\in \Tcal$.
Given any smooth vector field $X$, which no longer needs to be compactly supported, we write   
$$
\aligned
& \int_{e_K^+}  T_\alpha^\beta \, X^\alpha n_{K, \beta} \, dV_{e_K^+}
\\
& = \int_{e_K^-}  T_\alpha^\beta \, X^\alpha n_{K, \beta} \, dV_{e_K^+}
-
\sum \limits_{e^0\in \del^0 K}^{}\int_{e^0} T_\alpha^\beta \, X^\alpha n_{K,\beta} \, dV_{e_K^0}
+ \int_K \pi^{(X)}_{\alpha \beta}  T^{\alpha \beta}  dV_g,
\endaligned
$$
in which obvious notation has been used for the induced volume form along each boundary component of $\Kb$.
The approximation scheme is now defined from this general identity, by choosing $X$ to be either the vector $\del_t$ associated with the time function,
or vector fields tangent to the spacelike hypersurfaces.

Under our assumptions, the normal $n_{K, \beta }$ along the spacelike sides has components $(- N,0,0,0)$, hence the above equation becomes
$$
\aligned
& \int_{e_K^+}  T^{0\alpha} \, X_\alpha \, N\,dV_{e_K^+}
\\
& = \int_{e_K^-}  T^{0\alpha} \, X_\alpha \, N\,dV_{e_K^+}
- 
\sum \limits_{e^0\in \del^0 K}^{}\int_{e^0} T_\alpha^\beta \, X^\alpha n_{K,\beta} \, dV_{e_K^0}
+ \int_K \pi^{(X)}_{\alpha \beta}  T^{\alpha \beta}  dV_g.
\endaligned
$$
Finally, in specifically chosen coordinates, we can choose covector fields with constant components, say $X^{(0)} = (1, 0, \ldots)$, 
$X^{(1)} = (0,1, 0, \ldots)$, etc., and we can introduce the source-terms 
$$
S^\alpha = \pi^{(X^{(\alpha)})}_{\beta\gamma}  T^{\beta\gamma},
$$
which allows us to express the following averaged balance laws (for $\alpha= 0, 1, \ldots$)
\be
\label{eq:intform}
\int_{e_K^+}  T^{0\alpha} \, N\,dV_{e_K^+}
= \int_{e_K^-}  T^{0\alpha}  \, N\,dV_{e_K^+}
- 
\sum \limits_{e^0\in \del^0 K}^{}\int_{e^0} T^{\alpha\beta} \, n_{K,\beta} \, dV_{e_K^0}
+ \int_K S^\alpha \, dV_g.  
\ee

The geometric version of the finite volume method is based on the integral identity \eqref{eq:intform}.
The solution is represented, at every discrete time $t_n$ and within each spacelike hypersurface (say, $\Hcal_{t_n}$) by constant states $\uT^{0\bullet}_{e_K^-} = \big( \uT^{0\alpha}_{e_K^-} \big)$. The constant states $\uT^{0\bullet}_{e_K^+}$ on the ``next'' hypersurface (that is, $\Hcal_{t_{n+1}}$) are then determined as follows. Along each vertical face $e^0_K$ of an element $K$, we a priori fix a family of numerical flux functions, say, $F_{e_K^0}^\alpha \big( \uT^{0\bullet}_{e_K^-}, \uT^{0\bullet}_{e_{K'}^-}\big)$ (with $K'$ defined by $K \cap K' = e_K^0$ and $K \neq K'$), for the approximation of the vertical contribution, so that 
$$
T_\alpha^\beta \, n_{K,\beta} \simeq F^\alpha_{e_K^0}\big( \uT^{0\bullet}_{e_K^-}, \uT^{0\bullet}_{e_{K'}^-}\big). 
$$ 
The numerical fluxes are assumed to be locally Lipschitz continuous and satisfy standard properties of consistency and conservation~\cite{AmorimLeFlochOkutmustur}. The finite volume scheme generates the constants values $\uT^{0\bullet}_{e_K^+}$ and reads (for $\alpha= 0, 1, \ldots$)
$$
\uN_{e_K^+} \, |e_K^+| \,\uT^{0\alpha}_{e_K^+}
=
\uN_{e_K^-} \, |e_K^-| \, \uT^{0\alpha}_{e_K^-}
-
\sum_{e_K^0\in \del^0 K} |e_K^0| \, F_{e_K^0} \big( \uT^{0\bullet}_{e_K^-}, \uT^{0\bullet}_{e_{K'}^-}\big)
+ |K| \,  \Sb^\alpha,
$$
in which $\uN_{e_K^\pm}$ are consistent approximations of the lapse function on the corresponding hypersurfaces and $\Sb^\alpha$ are consistent approximations of the source--terms. 
For instance, under the symmetry assumption of main interest in this paper, we can work in the quotient manifold which has spacetime dimension $(1+1)$, and we need to introduce consistent approximations of  
$\Sb^0 = \Gamma_{00}^0 \uT^{00}  + \Gamma^0_{01} \uT^{01}$
and $\Sb^1 = \Gamma^1_{01} \uT^{01} + \Gamma_{11}^1 \uT^{11}$. 

Furthermore, a restriction on the time step is required for stability purposes. 
An extension of this scheme will be introduced in Section~4 below in order to make it to preserve steady state solutions at the discrete level. 
 

\section{Steady fluid flows on Schwarzschild spacetime}   
\label{sec:3}

\subsection{Steady solutions on a curved spacetime}

Our first task is to investigate the properties of steady state solutions to the Euler equations \eqref{ecur}. Without imposing symmetry assumptions, it is clear that no analytical closed formula could be derived for the solutions to 
\be
\label{ecur-sanst}
\aligned
&  \del_j \Big(BN^3\, (\rho c^2 + p) \, u^0 \, u^j \Big) 
   = S^0, 
\\
&  \del_j \Big(BN^3\Big(  (\rho c^2 + p) \, u^i \, u^j  +  p \, g^{ij} \Big) \Big) = S^i, 
\endaligned
\ee 
which is a mixed type (hyperbolic, elliptic) system in the variables $(x^j)$. By imposing spherical symmetry (for instance), this system reduces to a system of two ordinary differential equations, which, in itself is already quite chalenging. 
\be
\label{ecur-sanst1}
\aligned
&  \del_1 \Big(BN^3\, (\rho c^2 + p) \, u^0 \, u^1 \Big) 
   = S^0, 
\\
&  \del_1 \Big(BN^3\big(  (\rho c^2 + p) \, (u^1) ^2  +  p \, g^{11} \big) \Big) = S^1. 
\endaligned
\ee 
Furthermore, we point out that the solutions studied in the present section were first introduced in the physics literature in a different gauge 
\cite{HSW} and 
represent the steady state accretion of matter in a Schwarzschild black hole geometry. 


\subsection{Euler equations on Schwarzschild spacetime}

We are now interested in the outer domain of communication of Schwarz\-s\-child spacetime, 
which describes the exterior of a spherically symmetric black hole. The Schwarzschild metric is a particular solution to the Einstein equations and, in the so-called Bondi coordinates $(t,r, \theta, \varphi)$, reads   
\be
\label{schw}
ds^2 = -\Bigl( 1 - {2m \over r} \Bigr) c^2 \, dt^2 
 + \Bigl( 1 - {2m \over r} \Bigr)^{-1} \, dr^2
 + r^2 \bigl( d \theta^2 + \sin ^2 \theta \, d\varphi^2 \bigr),
\ee
which is meaningful for $r > 2 m$. The coefficient $m$
represents the mass of a black hole located at $r=0$. 
This spacetime is spherically symmetric, that is, is invariant under the group of 
rotations acting on the spacelike $2$-spheres of constant $t$ and $r$. 
It is static, since the vector field $\del_0=\del_t$ is a timelike Killing vector and asymptotically converges to the (asymptotically flat) Minkowski spacetime, when $r \to +\infty$. The expression \eqref{schw} represents the outer domain of communication, only, and a different coordinate choice would be required to go through the horizon located the ``boundary of the coordinates'' $r=2m$ (around which the spacetime itself is actually regular). 
 
From now on, for simplicity in the presentation we assume that the equation of state is linear, i.e. 
\be
\label{eq:400} 
p(\rho)=\sigma^2\rho, 
\ee
where $\sigma$ is a constant. This is not an essential assumption, however. 
 We restrict attention to solutions depending on the radial variable $r$, only, and such 
that the non-radial component of the velocity vanishes. In orther words, we have 
$(u^\alpha) = \big( u^0(t,r), u^1(t,r), 0,0 \big)$ and, as a consequence, the energy momentum tensor satisfies $T^{02}=T^{03}=T^{12}=T^{13}=T^{23}=0$. 

The velocity $u$ is a unit vector, thus  
$-1 = -  \bigl(1 - {2m \over r}\bigr) \, (u^0)^2 + \bigl(1 - {2m \over r}\bigr)^{-1}  (u^1)^2$ 
and, in terms of the rescaled velocity component $V =  {c \over 1 - {2m \over r}} \, {u^1 \over u^0}$, we find 
$$
(u^0)^2 = {c^2 \over (c^2-V^2) \bigl(1 - {2m \over r}\bigr) }, \qquad
(u^1)^2 = V^2 {\bigl(1 - {2m \over r}\bigr) \over (c^2 - V^2)}. 
$$
We can thus express the Euler system in the form  
$$
\aligned
&\del_0({BN^3}T^{0 0})+\del_r({BN^3}T^{1 0})=S^0,
\\
& \del_0 (BN^3T^{0 1})+\del_r (BN^3T^{11}) =S^1,
\endaligned
$$
in which  
$$
\aligned
S^0 &=0,\quad BN^3= \sin \theta\, r(r-2m),
\\
S^1&=T^{00}{BN^3\over 2}g^{11}\del_rN^2-BN^3\Gamma^1_{jj}T^{jj}+2N^3T^{11}\del_r B.
\endaligned
$$
We thus arrive at  
$$
\aligned 
& \del_t\left( r(r-2m) T^{00}\right)  + \del_r\left( r(r-2m)T^{01}\right)  = 0,
\\
& \del_t\left( r(r-2m)T^{01}\right)  + \del_r\left( r(r-2m)T^{11}\right) 
- 3mT^{11}
+ {c^2 m \over r^2}\left( r-2m\right)^2 \, T^{00} 
\\
& \hskip2.8cm -  r \left(r-2m \right)^2 \, T^{22} - r \, \sin^2 (\theta) \left(r-2m \right)^2 \, T^{33}
= 0,
\endaligned
$$
whose coefficients are given by  
\be
\label{eq:666}
\aligned
& \tT^{00} =    \bigl( 1 - {2m \over r} \bigr)T^{00}={c^2 \rho + p(\rho) V^2/c^2 \over c^2 - V^2} c^2,   
\qquad 
  \tT^{01} =T^{01}= {c^2 \rho + p(\rho) \over c^2 - V^2} cV,  
\\
& \tT^{11} = {1\over \Big(1 - {2m \over r} \Big)  }\,T^{11}={V^2 \rho + p(\rho) \over c^2 - V^2} c^2, 
\endaligned 
\ee
and 
$$
T^{22} = {p(\rho) \over r^2},\qquad T^{33} = {p(\rho) \over r^2 \, \sin^2 \theta}.
$$
In conclusion, the {\sl Euler system on a Schwarzschild background} takes the form
\be
\label{17h9} 
\aligned 
& \del_t\big({r^2\over c^2} \tT^{00}\big)  + \del_r\left( {r(r-2m)\over c}\tT^{01}\right)  = 0,
\\
& \del_t\left( {r(r-2m)\over c}\tT^{01}\right)  + \del_r\left( (r-2m)^2\tT^{11}\right) 
- 3m\,{r-2m\over r}\tT^{11}
\\
& \hskip2.9cm + m{r-2m\over r} \tT^{00} 
-  {2\sigma^2\over r}(r-2m)^2\,{\tT^{00}-\tT^{11}\over c^2-\sigma^2}
= 0.
\endaligned
\ee 

For later use, we record here some additional formulas:  
\be
\label{abs}
\aligned
&c^2T^{00}(1-{2m\over r})={r\over r-2m}T^{11}+\rho c^2-p,
\\
&\rho=\tT^{00}\,{1-V^2\over 1+\sigma^2V^2},
\\
&
V={1\over 2\sigma^2}\,{\tT^{00}\over\tT^{01}}\,\Bigg(1+\sigma^2-\sqrt{(1+\sigma^2)^2-4\sigma^2\,\bigg({\tT^{01}\over\tT^{00}}\bigg)^2}\Bigg),
\endaligned
\ee
and  
\be
\aligned
\label{iden1}
&\tT^{00}\,\tT^{11}-(\tT^{01})^2=\sigma^2\,c^2\,\rho^2,
\qquad 
\quad \tT^{00}+\tT^{11}={(\sigma^2+c^2)+(V^2+c^2)\over c^2-V^2 }, 
\\
&\tT^{00}-\tT^{11}=(c^2-\sigma^2)\,\rho,
\qquad \quad
(\tT^{11})^2-(\tT^{01})^2={\sigma^4-V^2c^2\over c^2-V^2}\,\rho^2c^2. 
\endaligned
\ee


\subsection{A closed formula for steady fluid solutions} 

We now derive a closed formula for smooth steady solutions $\rho=\rho(r)$ and 
$V=V(r)$ to the Euler equations posed on a Schwarzschild spacetime. 
Various plots of solutions are provided in Figures 3.1 to 3.4. 
We work in an interval $r \in (2m,R]$ for some fixed $R>2m$, and we impose boundary data at $r=R$
$$
\rho (R)>0, \qquad V(R) \in (-1,1). 
$$ 
The Euler equations reduce to the following system of ordinary differential equations: 
\be  
\label{ode1}
{d \over dr} \left( {r(r-2m)\over c} \, \tT^{01}\right)  = 0,
\ee
\be
\aligned
&  {d \over dr} \left( r(r-2m)\tT^{11}\right) 
- m\tT^{11}
+ m\tT^{00} 
-  2\sigma^2(r-2m)\,{\tT^{00}-\tT^{11}\over c^2-\sigma^2}
= 0.
\label{ode2}
\endaligned
\ee
The ``first'' equation \eqref{ode1}, after integration over the interval $[r,R]$, yields 
\be
\label{abf}
A(r) = r(r-2m) \, {c^2 \rho + p(\rho(r)) \over c^2 - V(r)^2} \, V(r) = A(R), \qquad \quad r \in (2m, R), 
\ee
where $A(R)$ is determined by boundary data prescribed at $r=R$. 
This implies that if $V(R) \gtrless 0$, then $V(r) \gtrless0$ for all $r \in (2m,R)$. By solving \eqref{abf} in terms of $V(r)$, we deduce 
\be
\label{ele}
V={-\kappa \, \rho\pm\sqrt{(\kappa \, \rho)^2+4c^2A^2}\over 2A},
\ee
where $\kappa = \kappa(r)=r(r-2m)(c^2+\sigma^2)$ is a function determined by the mass and the sound speed. 

By using our expressions \eqref{iden1}, the equation \eqref{ode1} (after taking the square) is equivalent to 
$$
{d \over dr} \left( {r^2(r-2m)^2\over c^2}\bigg(\tT^{00}\,\tT^{11}-{\sigma^2\,c^2\over (c^2-\sigma^2)^2}(\tT^{00}-\tT^{11})^2\bigg)\right)  
= 0.
$$
Hence, by introducing the following suitably weighted quantities
$$
\hatT^{00}=r(r-2m)\tT^{00},\quad \hatT^{11}=r(r-2m)\tT^{11},\quad \epsilon={\sigma\,c\over c^2-\sigma^2},\quad
 \mu={2\sigma\over c},
$$
the Euler system becomes 
\be  
\label{ode4}
{d \over dr} \left( {1\over c^2}\big(\hatT^{00}\,\hatT^{11}-\epsilon^2(\hatT^{00}-\hatT^{11})^2\big)\right)  = 0,
\ee
\be
\aligned
&  {d \over dr} \hatT^{11}+{1\over r}\,\bigg({m\over r-2m}-\mu\,\epsilon\bigg)\,(\hatT^{00}-\hatT^{11})
= 0.
\label{ode5}
\endaligned
\ee

We can integrate the equation \eqref{ode4}, as we did earlier, and with the new notation we now have 
$$
\hatT^{00}\,\hatT^{11}-\epsilon^2(\hatT^{00}-\hatT^{11})^2=c^2 \, A^2,
$$
where we recall that $A=A(R)=A(r)$ is a constant. 
By setting $Y=\hatT^{11}$, this equation takes the form
$$
-\epsilon^2(\hatT^{00}-Y)^2 + \hatT^{00} Y- c^2 \, A^2=0,
$$
which leads us to 
\be
\label{zero}
\hatT^{00}=Y\,\bigg({1\over 2\epsilon^2}+1\bigg) \mp {1\over 2\epsilon^2} \sqrt{Y^2(1 +4\epsilon^2)-4\epsilon^2c^2A^2}. 
\ee
In addition, since the dominant energy condition (satisfied by the fluids under consideration in \eqref{eq:400}) 
imposes $T^{00}\geq T^{11}$, we must have $\hatT^{00}-Y\geq0$, that is, 
$$
Y \mp \sqrt{Y^2(1 +4\epsilon^2)-4\epsilon^2c^2A^2}\geq0,
$$
which is a constraint on the values taken by $Y$. 
Using the definition of $Y$ and the expression \eqref{zero} in the ``second'' Euler equation \eqref{ode5}, we now find 
\be  
{dY \over dr} + {1\over r}\,\bigg({m\over r-2m}-\mu\,\epsilon\bigg)\,{1\over 2\epsilon^2}\bigg(Y \mp \sqrt{Y^2(1 +4\epsilon^2)-4\epsilon^2c^2A^2}\bigg)
= 0.
\ee
This differential equation in $Y$ can be solved and we obtain 
$$
  {dY\over Y \mp \sqrt{Y^2(1 +4\epsilon^2)-4\epsilon^2c^2A^2}}=-{1\over r}\,\bigg({m\over r-2m}-\mu\,\epsilon\bigg)\,{1\over 2\epsilon^2}\,dr 
$$
and, after integration, 
$$
\aligned
\int_r^R {dY\over Y \mp \sqrt{Y^2(1 +4\epsilon^2)-4\epsilon^2c^2A^2}}
&
={1\over 4\epsilon^2}\,\int_r^R\bigg({1+2\mu\,\epsilon\over r}-{1\over (r-2m)}\bigg)\,dr
\\
&=\ln \Bigg(\Big({R \over r}\Big)^{1+2\mu\epsilon\over 4\epsilon^2 }\Big({r-2m\over R-2m}\Big)^{1\over 4\epsilon^2}\Bigg)\\
&
  =\ln \Bigg( {F(r) \over F(R)} \Bigg). 
\endaligned
$$

  \begin{figure}[htp]
  \begin{center}
  {\includegraphics[height=7.6cm,angle=90]{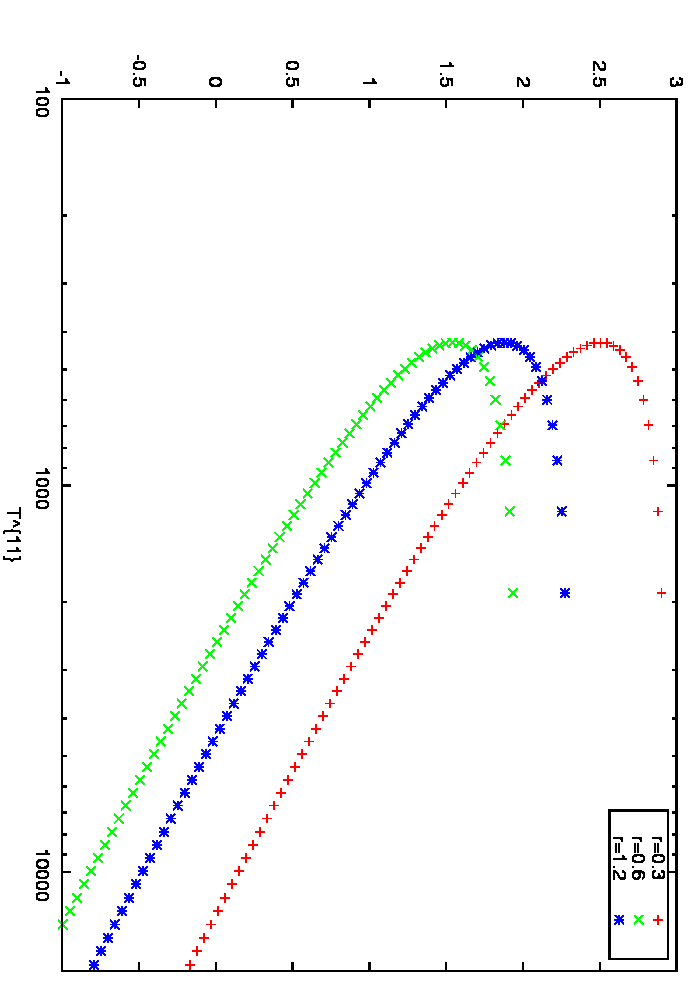}}
\end{center} 
  \begin{center}
  {{\bf Figure 3.1.} Steady solutions for three values of the radius $r$.}
\end{center}  
  \end{figure}

  \begin{figure}[htp]
  \begin{center}
  {\includegraphics[height=6.2cm,angle=-0]{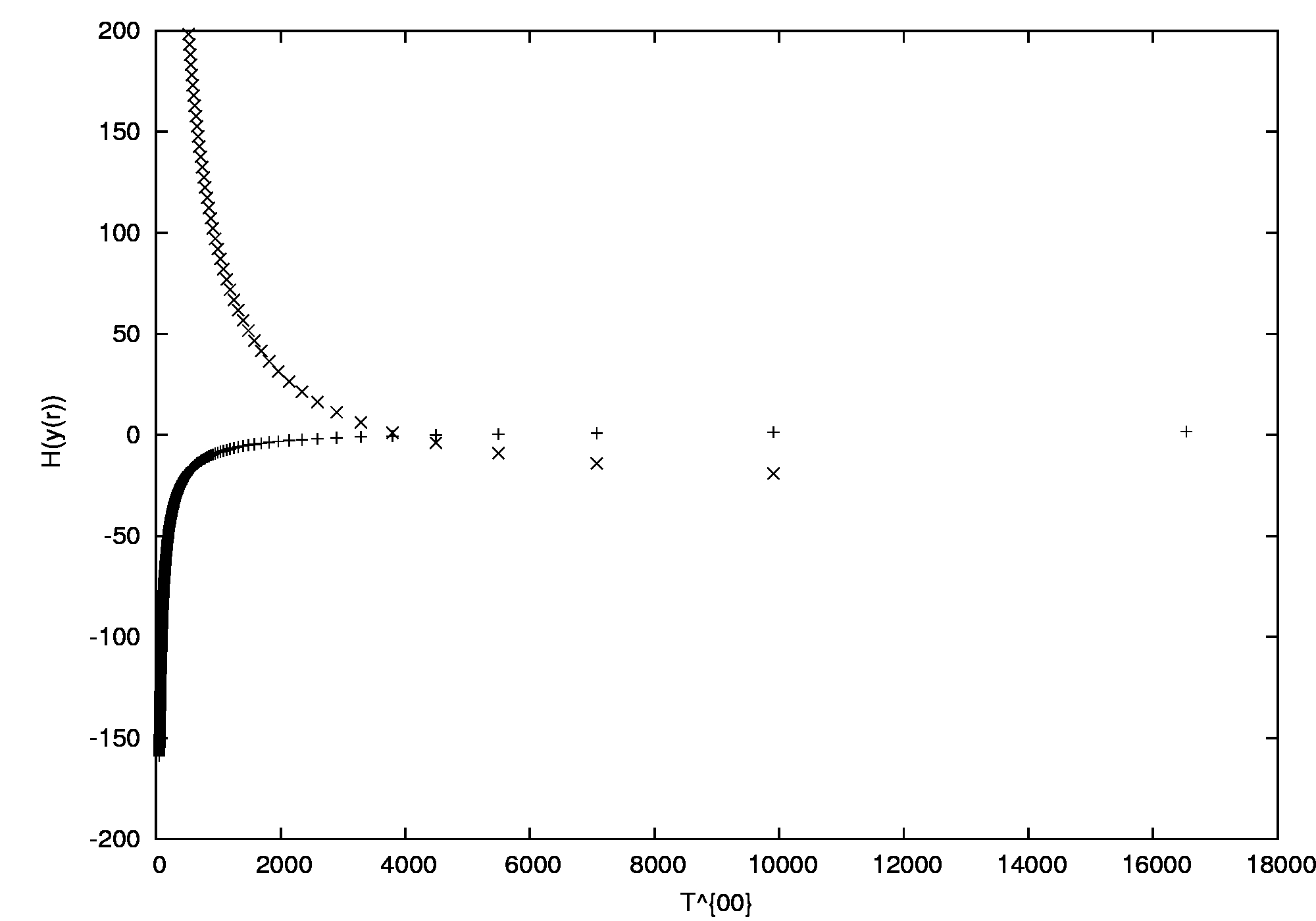}}
  \end{center}
\begin{center}
  {{\bf Figure 3.2.} Two steady solutions $Y=\tT^{11}$ for the radius $r=.8$ 
\newline 
with $\rho(R)=460$ and $V(R)=.001$.}
\end{center} 
  \end{figure}

\begin{figure}[htp]
  \begin{center}
  {\includegraphics[height=5cm,angle=0]{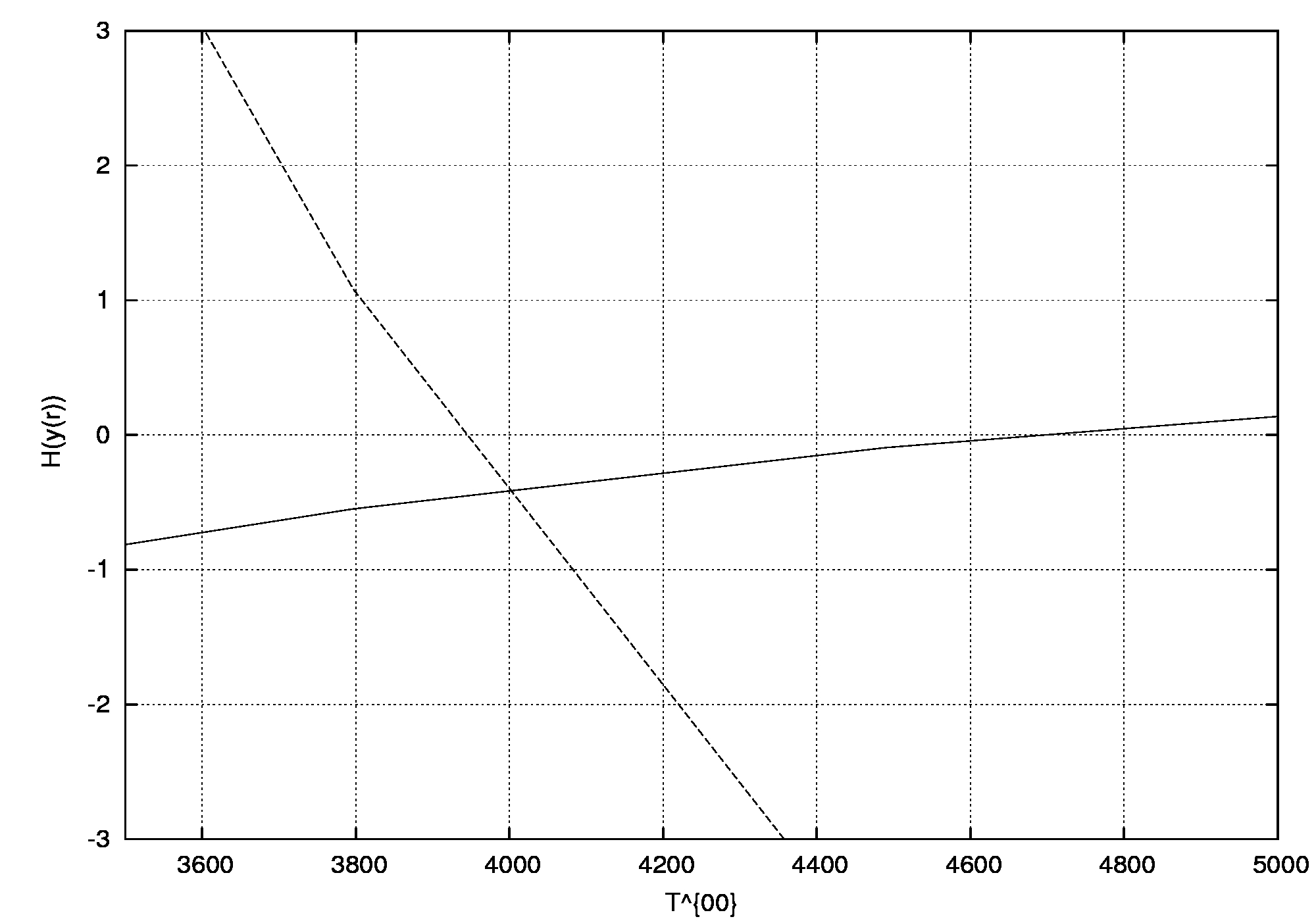}}
  \end{center}
\begin{center}
  {{\bf Figure 3.3.} Two steady solutions $\tT^{11}$ for a fixed radius $r=.8$
\newline 
 with $\rho(R)=460$ and $V(R)=.001$.}
\end{center}
  \end{figure}

With the change of variable $y^2 = Y^2\,\Lambda^2 = Y^2 {1+4\epsilon^2\over 4c^2A^2\epsilon^2}$, the integrand simplifies:  
$$
\aligned
\ln \Big( {F(r) \over F(R)}\Big)
& = 
\int_r^R {dy\over y \mp \delta\sqrt{y^2-1}}
\\
& = 
\int_r^R {d(\cosh\,X(y))\over \cosh\,X(y)\mp \delta\,\sinh\,X(y)}
=
 {1\over (1\mp\delta)}\int_r^R {e^{2X(y)}-1\over e^{2X(y)}+\zeta} \, dX,
\endaligned
$$
where we have used the change of variable $y=\cosh\,X(y)$ and set $\delta=(1+4\epsilon^2)^{1/2}$
and $\zeta={1\pm\delta\over 1\mp\delta}$. We thus obtain 
$$
\aligned
\ln \Big( {F(r) \over F(R)}\Big)
& = 
 {1\over (1\mp\delta)}\bigg(-X(y(r))+X(y(R))\bigg)-{\zeta+1\over 2(1\mp\delta)}\int_r^R {1\over e^{2X(y)}+\zeta}\,dX
\\
& =  {1\over (1\mp\delta)}\bigg(-X(y(r))+X(y(R))\bigg)-{\zeta +1 \over 2(1\mp\delta)}\int_r^R {1\over W( W+\zeta)}\,dW
\endaligned
$$
with $e^{2X(y)}=W$ and, so by integration, 
$$
\aligned
\ln \Big( {F(r) \over F(R)}\Big)
&={-1\over 1\mp\delta}\bigg(X(y(r)) - X(y(R))\bigg)-{1\over 1-\delta^2 }\Bigg(\ln{W(R)\over W(r)}-\ln{W(R)+\zeta\over W(r)+\zeta}\Bigg) 
\\
 &={-1\over 1\mp\delta}\bigg(X(y(r)) - X(y(R))\bigg)
\\
& \hskip1.cm 
     + {1\over 1-\delta^2 }\Bigg(2 \big( X(y(r)) - X(y(R)) \big) +\ln{e^{2X(y(R))}+\zeta\over e^{2X(y(r))}+\zeta}\Bigg) 
\\
 &=\bigg(-{1\over 1\mp\delta}+{2\over 1-\delta^2 }\bigg)\bigg(X(y(r))-X(y(R))\bigg)+{1\over 1-\delta^2 }
\Bigg(\ln{e^{2X(y(R))}+{1\pm\delta\over 1\mp\delta}\over e^{2X(y(r))}+{1\pm\delta\over 1\mp\delta}}\Bigg).
\endaligned
$$

\begin{figure}[t!]
\centering \subfloat[]{\label{a14}
    \begin{minipage}[b]{0.48\linewidth}
        \centering \includegraphics[width=0.98\linewidth]{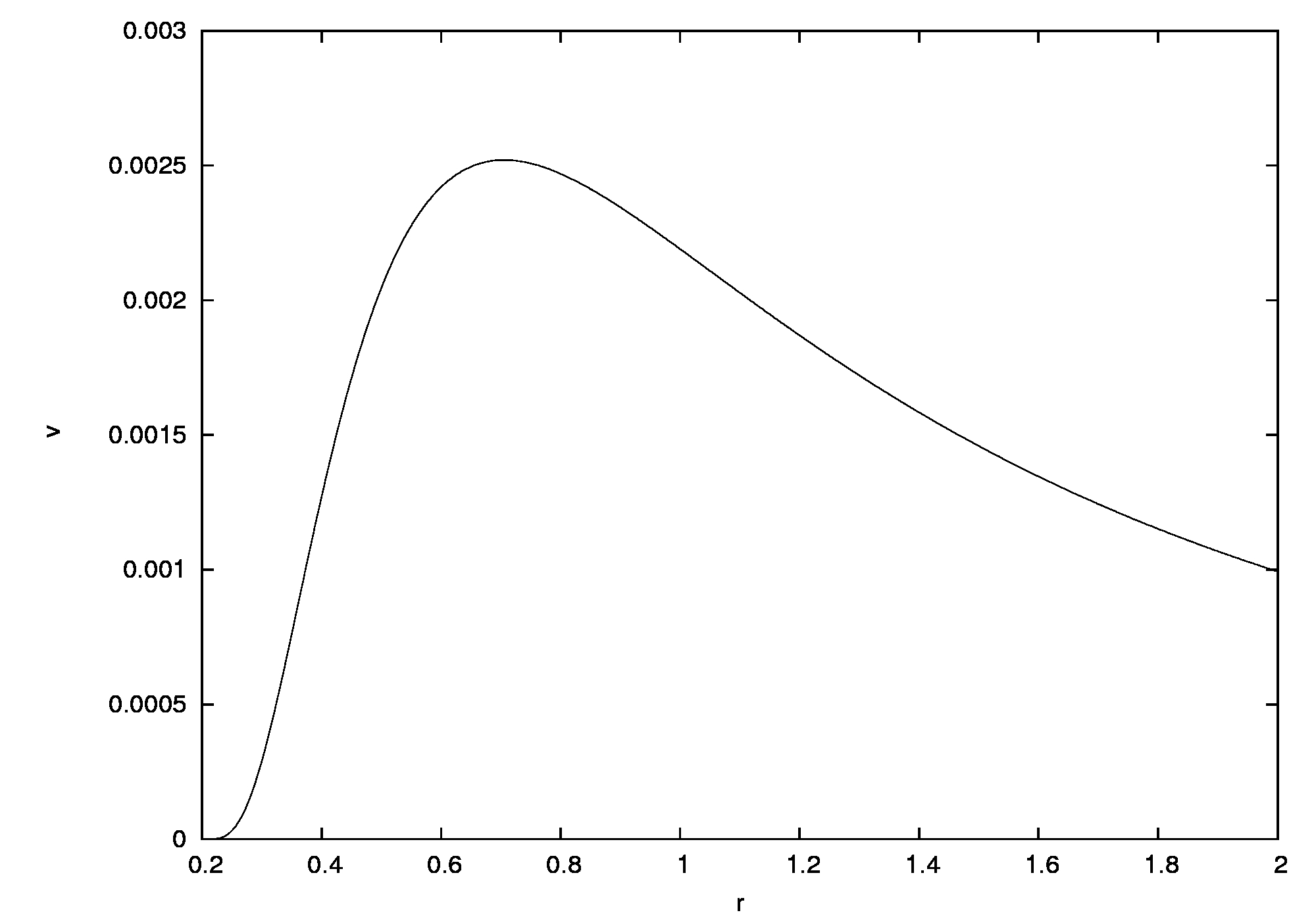}
    \end{minipage}}
\subfloat[]{\label{a24}
    \begin{minipage}[b]{0.48\linewidth}
        \centering \includegraphics[width=0.98\linewidth]{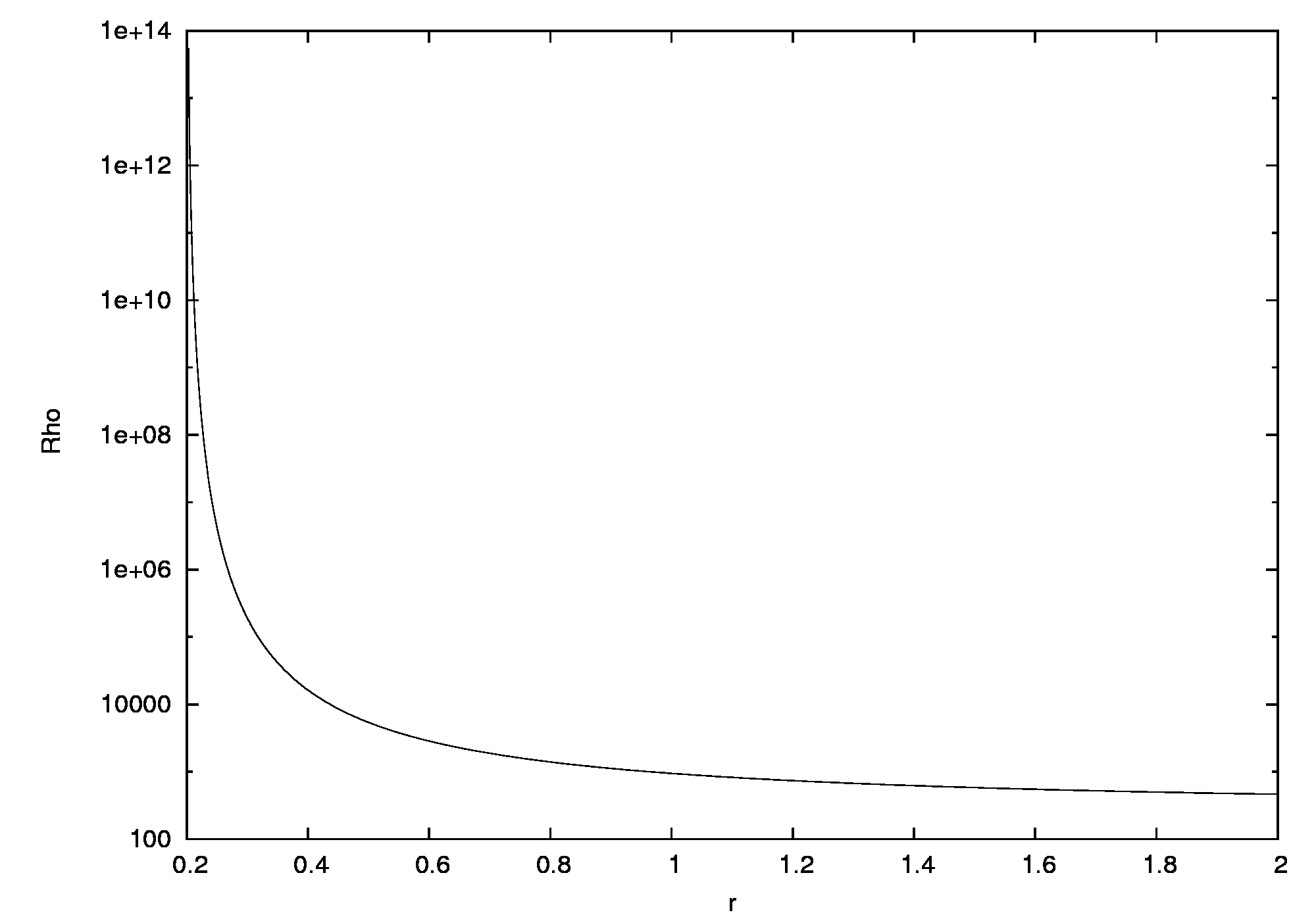}
    \end{minipage}
     }
\begin{center}
{{\bf Figure 3.4.} Steady solutions on Schwarzschild spacetime: (a) velocity \,  (b) density
\newline 
with 
$\rho(R)=460$, $V(R)=0.001$, $\sigma=0.3$, $R=2$, $2m=0.2$.}
\end{center}
\end{figure}


Therefore, we have derived an algebraic relation for the function $X=X(y(r))$, that is, 
\be
\label{summ}
\aligned
& H(y(r))
 = \Bigg({ y(r)+\sqrt{y(r)^2-1}\over y(R)+\sqrt{y(R)^2-1} }\Bigg)^{1\mp\delta}
\Bigg({y(R)+\sqrt{y(R)^2-1}+{1\pm\delta\over 1\mp\delta}\over y(r)+\sqrt{y(r)^2-1}+{1\pm\delta\over 1\mp\delta}}\Bigg)
-
\Bigg( {F(r)\over F(R)} \Bigg)^{1-\delta^2}
\\
& =0,
\endaligned
\ee  
where we recall that 
$$
\aligned
&\epsilon={\sigma\,c\over c^2-\sigma^2},\quad\delta=(1+4\epsilon^2)^{1/2},\quad\zeta={1\pm\delta\over 1\mp\delta},
\quad
F(r)= \bigg({1 \over r}\bigg)^{1+2\mu\epsilon\over 4\epsilon^2 }\bigg({r-2m}\bigg)^{1\over 4\epsilon^2},
\\
&\Lambda^2={1+4\epsilon^2\over 4c^2A^2\epsilon^2},
\quad Y= r(r-2m)\,T^{11},\quad y^2=Y^2\,\Lambda^2.
\endaligned
$$
 
In summary, given $\rho(R) >0$ and $V(R) \in (-1,1)$, there exist two steady solutions 
$$
y_\pm=y_\pm(r), \qquad r\in (2m, R]
$$
to the reduced Euler equation \eqref{summ}. We have derived an algebraic relation for these solutions, specifically $H(y(r))=0$, so that 
$y(r)$ can be computed by solving this equation for {\sl each} $r$, for instance by a fixed point technique. 
From the function $y=y(r)$, it is then straightforward to recover the physical variables. Namely, by using \eqref{zero}, 
$$
\hatT^{00}=Y\,\bigg({1\over 2\epsilon^2}+1\bigg) \mp {1\over 2\epsilon^2} \sqrt{Y^2(1 +4\epsilon^2)-4\epsilon^2c^2A^2}, 
$$
and $\tT^{00}-\tT^{11}=(c^2-\sigma^2)\,\rho$, 
we obtain the expression of the density $\rho=\rho(r)$ and, next, for the velocity $V=V(r)$ from \eqref{ele}, that is,
$$
V ={- \kappa \, \rho\pm\sqrt{\kappa^2 \, \rho^2+4c^2A(R)^2}\over 2A(R)}.
$$


\section [Well--balanced approximations]{Well--balanced approximations on Schwarzschild spacetime}   
\label{sec:4} 

\subsection{Finite volume method}

We express the Euler equations on Schwarzschild spacetime in the form of a hyperbolic system of balance laws, that is, 
\be
\label{bl}
\del_t\Up+\del_r\Fp(\Up, r)=\Sp(\Up,r), \qquad r > 2m, 
\ee
with, in view of \eqref{eq:666}-\eqref{17h9},  
$$
\Up=\begin{pmatrix}U^0\\U^1\end{pmatrix}=
\begin{pmatrix}r^2\,{c^2 \rho + p(\rho) V^2/c^2 \over c^2 - V^2}  \\r(r-2m)\,{c^2 \rho + p(\rho) \over c^2 - V^2} \,V\end{pmatrix},
$$
$$
\Fp(\Up, r)=\begin{pmatrix} F^0(\Up, r)\\F^1(\Up, r)\end{pmatrix}
=
\begin{pmatrix} r(r-2m)\,{c^2 \rho + p(\rho) \over c^2 - V^2} \,V\ \\(r-2m)^2{V^2 \rho + p(\rho) \over c^2 - V^2}\,c^2\end{pmatrix},
$$
and
$$
\aligned
\Sp(\Up,r)
& =\begin{pmatrix}S^0(\Up, r) \\ S^1(\Up, r)
\end{pmatrix}
\\
& =\begin{pmatrix}0\\ 
{r-2m\over r \, (c^2 - V^2)} \rho \, \Big(
\big(3 m (c^2 + \sigma^2) - 2 \sigma^2 r\big) \, V^2 
- c^2 \big( m (c^2+ \sigma^2) - 2 \sigma^2 r \big)\Big)
\end{pmatrix}.
\endaligned
$$

We apply the finite volume technique, presented in Section~\ref{sec:volfini} for general systems of balance laws, by 
working here with the $(1+1)$--dimensional quotient (by the group of spatial symmetries) of the Schwarzschild spacetime. For this quotient metric $g$, 
we thus have $ds^2 = -c^2\,\bigl( 1 - {2m \over r} \bigr) \,dt^2 + \widetilde{g}$ with
 induced volume form 
$$
dV_{\widetilde{g}_t}= (\widetilde{g})^{1/2} \,dr 
= \Big( 1-{2m\over r} \Big)^{-1/2} \,dr, 
\qquad \quad 
\Delta  V_{\gt_t}=\int_{r_{j-1/2}}^{r_{j+1/2}}  \Big( 1-{2m\over r} \Big)^{-1/2}  \, dr.
$$  
In agreement with Section~\ref{sec:volfini} and by introducing the approximations  
$$
\Ur_j^n  \simeq {1\over \Delta r_j}\int_{r_{j-1/2}}^{r_{j+1/2}}\Up(r,t_n)\,dV_{{\gt_{t_n}}},
\qquad\quad
\Sr_{j}^n \simeq {1\over \Delta r_j}\int_{r_{j-1/2}}^{r_{j+1/2}}\Sp(t_n,r)\,dV_{{\gt_t}}, 
$$
and  
$$
\Delta  r_j = \Delta V_{\gt_{t_n}}=\int_{r_{j-1/2}}^{r_{j+1/2}} \Big( 1-{2m\over r} \Big)^{-1/2}  \, dr,
$$ 
the finite volume scheme takes the form 
\be
\label{full}
\Ur^{n+1}_{j}=\Ur^n_{j}-{\Delta  t\over \Delta  r_j}\big(\Fr_{j+1/2}^n - \Fr_{j-1/2}^n \big)+\Delta  t\, \Sr_{j}^n, 
\ee 
in which $\Fr_{j+1/2}^n$ are consistent approximations of the exact flux of the system \eqref{bl}. 

As usual, the Courant--Friedrichs--Lewy (CFL) condition  is imposed on the time step in order to guarantee stability. Specifically, 
we set $\Delta t= t_{n+1} - t_n$, which we assume to be independent of $n$ for simplicity, and impose the inequality 
\be
\label{eq:501} 
{\Delta  t\over \Delta  r} \, \max | \lambda(\Up)| < 1, 
\ee
where the maximum is taken over the (real)  wave speeds $\lambda(\Up)$ of the Euler system. 


\subsection{Taking the Schwarzschild geometry into account}
 
In steady state solutions to \eqref{bl}, which by definition satisfy 
\be
\label{stea}
\del_r\Fp(\Up, r)=\Sp(\Up,r), 
\ee
the source terms {\sl exactly balance} the flux terms. We will construct the well-balanced version of the finite volume scheme introduced in the previous subsection by imposing that the same property must hold at the discrete level of approximation for the family of discrete steady states. 
For instance, {\sl cell--centered} evaluation of the source terms, generally, do not ensure the preservation of these discrete steady states. Therefore, we look for an adapted discretization of the source-term which directly uses information from the steady state equation and, more specifically, uses the characterization \eqref{summ} exhibited in Section~3, above.  In turn, our scheme will satisfy a discrete version of the steady state system \eqref{stea}. 

Motivated by the work by Russo et al.~\cite{PuppoRusso,Russo1}, the reconstruction scheme proposed now takes 
 the family of steady solutions into account for the {\sl numerical evaluation of the intermediate states}
 at which the numerical flux is computed. 
Specifically, recalling the first Euler equations (in the form adopted i the present paper) contains no source-term, 
we define the well--balanced finite volume scheme   
\be
\label{DS1}
\aligned 
\Ubar_{j}^{0,n+1}
& = \Ubar_{j}^{0,n} - \frac{\Delta t}{\Delta r_j}\big( \Fbar_{j+1/2}^{0,n}
        - \Fbar_{j-1/2}^{0,n}\big), 
\\
\Ubar_{j}^{1,n+1}
& = \Ubar_{j}^{1,n} - \frac{\Delta t}{\Delta r_j}\big( \Fbar_{j+1/2}^{1,n}  - \Fbar_{j-1/2}^{1,n} \big)+ \Delta t \, \Sbar_j^{1,n}, 
\endaligned
\ee 
by substituting our closed expression of the family of steady solutions. Clearly, the first equation is in a conservation form and does not need any well-balanced correction, so we concentrate on the second equation. 

First of all, we approximate the solution in each cell by a steady solution and, under this approximation and by 
using integration by parts, we can then transform 
the source term for the second Euler equation (cf.~the expression \eqref{17h9}), as follows (for exact steady solutions):  
$$ 
\aligned
& {1\over \Delta  V_{\gt_t}}\,\int_{r_{j-1/2}}^{r_{j+1/2}} S_j^1\,dV_{\gt_t}
\\
&= {1\over \Delta  V_{\gt_t}}\,\int_{r_{j-1/2}}^{r_{j+1/2}}\Bigg(3m\,{r-2m\over r}\tT^{11}- m{r-2m\over r} \tT^{00} 
+{2\sigma^2\over r}(r-2m)^2\, {\tT^{00}-\tT^{11}\over c^2-\sigma^2} \Bigg)\,dV_{\gt_t}
\\
&={1\over \Delta r_j} \,\int_{r_{j-1/2}}^{r_{j+1/2}}\del_r\left( (r-2m)^2\tT^{11}\right)\,  \Big( 1-{2m\over r} \Big)^{-1/2}  \, dr 
\\
&={1\over \Delta r_j}\,\Big( r^{1/2}(r-2m)^{3/2}\tT^{11}\Big)\Big|_{r_{j-1/2+}}^{r_{j+1/2-}}
+
{m \over \Delta r_j}\int_{r_{j-1/2}}^{r_{j+1/2}} \tT^{11} \, dV_{\gt_t}. 
\endaligned 
$$ 
This identity, valid for exact steady solutions, motivates us to propose the following approximation for the source term
\be
\label{17h9-2} 
\aligned
\overline S_j^{1,n}
= \, 
&{1\over \Delta r_j}\,\Big( r^{1/2}(r-2m)^{3/2} \overtT^{11, n}_{j+1/2 -} - r^{1/2}(r-2m)^{3/2} \overtT^{11,n}_{j-1/2+} 
\Big)
\\
&
+
{m \over \Delta }\int_{r_{j-1/2+}}^{r_{j+1/2-}} \tT^{11,n} \,  \Big( 1-{2m\over r} \Big)^{-1/2}  \, dr. 
\endaligned 
\ee 
Of course, in order to be able to make use of the above definition, it remains, on one hand,
 to introduce suitable approximations $\overtT^{11,n}_{j+1/2-}$ and $\overtT^{11,n}_{j-1/2 +}$ 
(consistent with the values taken by the ``true'' solution) at 
the interfaces between the cells 
and, on the other hand, 
to evaluate the integral term above. 

Consider first the latter issue, we propose here to use Simpson's rule, and we replace the integral term by the following explicit expression 
\be
{m \over \Delta  V_{\gt_t}}\int_{r_{j-1/2+}}^{r_{j+1/2-}} \tT^{11,n} \, dV_{\gt_t} 
\approx 
{m  \over 6}\,\Big( \overtT^{11,n}_{j-1/2+} + 4 \, \overtT^{11,n}_{j} + \overtT^{11,n}_{j+1/2-}  \Big). 
\ee

Observe next that, since we have introduced new states at the interfaces between the cells, it is natural (and actually necessary in order to achieve the weel-balanced property) to compute the numerical flux (of the second Euler equation) in terms of these interfaces value. In other words, we write 
(at $j+1/2$, say, and with similar formulas at $j-1/2$) 
\be
\label{NFLR}
\aligned
\Fbar_{j+1/2}^0 
& =\Fbar^0\big(\Ubar_{j+1/2-}^1,\Ubar_{j+1/2+}^1\big), \quad
\qquad 
\\
\Fbar_{j+1/2}^1
& =\Fbar^1\big(\Ubar_{j+1/2-}^0,\Ubar_{j+1/2-}^1, \Ubar_{j+1/2+}^0, \Ubar_{j+1/2+}^1 \big), 
\endaligned
\ee
where we are taking in to account the particular dependency of the flux of the Euler system. 
By taking the Schwarzschild geometry into account, we can write (with obvious notation)
$$
\aligned
&\Ubar_{j+1/2-}^{1,n} = r_{j+1/2}(r_{j+1/2}-2m) \, \overtT^{01,n}_{j+1/2-}, 
\\
&
 \Ubar_{j+1/2+}^{1,n} =r_{j+1/2}(r_{j+1/2}-2m) \, \overtT^{01,n}_{j+1/2+}, 
\endaligned
$$

It now remains to compute the states at the interfaces. First of all, recalling that, for exact steady solutions, the expression $r(r-2m)T^{01}$ is a constant, we naturally determined 
the ``reconstructed'' states $\overline{\tT}^{01,n}_{j\pm1/2\mp}$ and $\overtT^{01,n}_{j\pm1/2\mp}$ at the interfaces 
(by interpolation from the states within the cells) by setting 
$$
\aligned
&\overtT^{01,n}_{j+1/2+} =  \, {(r_j+\Delta r_j)(r_j+\Delta r_j-2m) \over r_{j+1/2}(r_{j+1/2}-2m)} \, \overtT^{01,n}_{j+1}, 
\\ 
&\overtT^{01,n}_{j+1/2-} =  \, {r_j(r_j-2m) \over r_{j+1/2}(r_{j+1/2}-2m)} \, \overtT^{01,n}_{j},
\\
&\overtT^{01,n}_{j-1/2+} =  \, {r_j(r_j-2m) \over r_{j-1/2}(r_{j-1/2}-2m)} \, \overtT^{01,n}_{j},
\\
&\overtT^{01,n}_{j-1/2-} =  \, {(r_j-\Delta r_j)(r_j-\Delta r_j -2m) \over r_{j-1/2}(r_{j-1/2}-2m)} \, \overtT^{01,n}_{j-1}.
\endaligned
$$
On the other hand, the ``reconstruction" of the interface states $\overtT^{11,n}_{j\pm1/2\mp}$ and $\overtT^{11,n}_{j\pm1/2\mp}$ 
is more delicate and requires the full algebraic relation \eqref{summ}, which provides us with a complete characterization of steady solutions. This characterization is based on the function $H= H(y)$ discovered in Section~3 and, therefore, in each computational cell we now impose the two relations (with obvious notatopn for the quantities $\overline y_j^n$, $\overline{y}^n_{j-1/2+}$, and $\overline{y}^n_{j+1/2-}$) 
$$
\aligned 
\qquad
&H(\overline{y}^n_{j+1/2-})=H(\overline{y}^n_{j}) = H(\overline{y}^n_{j-1/2+}),  
\qquad 
\endaligned
$$
in which the value $\overline y_j^n$ is explicitly known from the states in the cell $j$, while 
 the unknowns $\overline{y}^n_{j-1/2+}$ and $\overline{y}^n_{j+ 1/2-}$ 
at the interfaces
are obtained by (numerically) solving the above algebraic equations (by a fixed point technique). 
Thanks to the relation \eqref{zero}, we can now determine all of the interface values $\overtT^{00,n}_{j\pm1/2\mp}$ and $\overtT^{00,n}_{j\pm1/2\mp}$. 
Finally, from the identity \eqref{iden1}, we can also compute the mass energy density states $\overline{\rho}^n_{j\pm1/2\mp}$ as well as the velocity components $\overline{V}^n_{j\pm1/2\mp},$ and $\overline{V}^n_{j\pm1/2\mp}$. We have now completed the design of our 
well-balanced scheme (at the first--order of accuracy). 

In summary, the above construction relying on steady solutions in order to define the interface states for the evaluation of the numerical fluxes, it can be checked that steady solutions are exactly preserved at the discrete level of approximation. On the other hand, the consistency property of the original scheme (ensuring that limits of the scheme do satisfy the Euler equations in the sense of distributions)
also holds.


\subsection{Second--order accuracy}

To arrive at a second--order scheme, we follow Nessyahu and Tadmor \cite{NessyahuTadmor} and introduce a predictor--corrector scheme, as follows:   
$$
\aligned
&\Ur^{n+1/2}_{j} = \Ur^{n}_{j} - {\Delta   t \over 2\Delta  r} \, \Fr_j', 
\\
&\Ur^{n+1}_{j+1/2}={1\over 2} \, \Big( \Ur^{n+1}_{j } + \Ur^{n}_{j+1} \Big)  \hskip-.05cm
+ {1\over 8}(\Ur'_{j}-\Ur'_{j+1})
-
{\Delta   t \over \Delta  r} \bigg(\Fr \big(\Ur^{n+1/2}_{j+1} \big) \hskip-.1cm -\Fr \big( \Ur^{n+1/2}_{j} \big)\bigg). 
\endaligned
$$ 
The states $\Ur_j',\Fr'_j$ represent first--order approximations of the space derivatives (of the field and flux variables) at the point $r_j$
and can be computed in several ways. A standard choice (which is used in this paper) is given by 
$$
\aligned
\Ur_{j}' &= \MM\big( \Ur^n_{j+1} - \Ur^n_{j},\,\Ur^n_{j} -\Ur^n_{j-1}  \big),
\\
\Fr'_j &= \MM \big(\Fr_{j+1}^n - \Fr_j^n,\, \Fr_j^n -\Fr_{j-1}^n \big),
\endaligned
$$
where $\MM(U, W)$ is the min-mod limiter (which we apply composent-wise)      
$$
\MM(U,W)= \begin{cases}
\sgn(U) \, \min(|U|, |W|),     \qquad     & \sgn(U) = \sgn(W),
\\ 
0,            & \text{ otherwise.}  
\end{cases} 
$$


\section{Numerical experiments and applications}
 \label{sec:5}

\subsection{Comparison between several schemes}

In the following numerical experiments, we investigate the proposed scheme for the computation of weak solutions to the Euler equations on Schwarzschild spacetime. We work within the exterior domain of communication $r\in (2m, R)$ limited by the horizon $r=2m$ and 
a sphere with radius $R> 2m$. In all tests, the sound speed is taken to be $\sigma =0.3$, the light speed is unit, the upper space bound $R=2$, and the mass parameter is $m=0.1$. More precisely, our computations take place in an interval $r \in (r_0, R)$, with $r_0>2m$, so that we stay away from the horizon, on which the Euler equations (in the chosen coordinates) are singular.  
In every test, we treat the boundaries at $r=r_0$ and $r=R$ by solving a Riemann problem between the boundary data 
(determined from the given initial data) and the current numerical values at the boundaries, and we use
the flux of the Riemann solutions, which is the Godunov scheme at the boundary~\cite{DuboisLeFloch,DuboisLeFloch2}.  

We begin by illustrating the interest of the well--balanced property and we compare together two schemes, the proposed well-balanced one as well as a ``naive discretization'' (cf.~next paragraph) of the right--hand sides of the Euler equations. Throughout, we also plot the exact (or asymptotic) solution when available. 
The ``naive discretization'' of the right--hand sides of the Euler equations is defined by replacing the discretiation of the source $\Sr_{j}^n$ in \eqref{full} by an evaluation computed from the state $\Ur^n_{j}$, that is, {\sl for the naive version, only,} $\Sr_{j}^n$ reads 
$$ 
\begin{pmatrix}0\\ 
{r_j-2m\over r _j \, (c^2 - (V_j^n )^2)} \rho_j^n \, \Big(
\big(3 m (c^2 + \sigma^2) - 2 \sigma^2 r_j \big) \, (V_j^n )^2 
- c^2 \big( m (c^2+ \sigma^2) - 2 \sigma^2 r_j \big)\Big)
\end{pmatrix}.
$$

The numerical results are plotted in Figures~5.1.1 and 5.1.2, where we have chosen an initial data made of a perturbation of a {\sl steady solution.} This solution is determined from the value of the velocity $V(R) =0.001$ and the density $\rho(R)= 460$
at the end point of the interval $(r_0, R)$. A sine perturbation is added to the steady solution, so that a genuine evolution in time now takes place and we work in the interval $r_0 = 0.5 < r < R= 2$. We use the space mesh size $\Delta r=0.025$ and a CFL number equal to $0.9$. 

Figure 5.1.1 (a) represents the steady solution (velocity component) together with its perburbation which serves as our initial data. 
Figure 5.1.1 (b) represents the steady solution together with the numerical solutions with our two schemes. Observe that the well-balanced scheme produces a solution which is closer to the steady solution and oscillates about it, while the standard scheme deviates significantly from it. 
Figure 5.1.1 (c) represents the same solutions, but at a much later time: we now observe that the well--balanced numerical solution slightly oscillates about the steady solution, while the standard numerical solution is clearly completely wrong. 
Figure 5.1.1 (d) represents the time--asymptotic behavior, and we observe that the well-balanced scheme has re-converged to the original steady solution, the perturbations having cancelled out asymptotically, while again the standard scheme has generated a completely wrong solution.

Figure 5.1.2, instead of the physical variables (like the velocity above), shows the nonlinear expression $r(r-2m)\,\tT^{01}$, which is known to be constant for steady solutions. Figure 5.1.2 (a) represents this function for the steady solution (which is thus a constant) and for the well--balanced scheme after $N= 500$ and $N= 1000$ iterations, respectively.  Figure 5.1.2 (b) is a plot of the relative error for the same quantity. Again, we observe that the well--balanced scheme produces a quite satisfactory result with $0.5\%$ at the end point $r=R$ of the interval. The accuracy is better near the horizon but grows with $r$.  (A further correction of the scheme may be found useful to improve the accuracy for large radius $r$.)

\begin{figure}[t!]
\centering \subfloat[Initial data and perturbation (velocity) 
]{\label{a15}
    \begin{minipage}[b]{0.48\linewidth}
        \centering \includegraphics[width=0.98\linewidth]{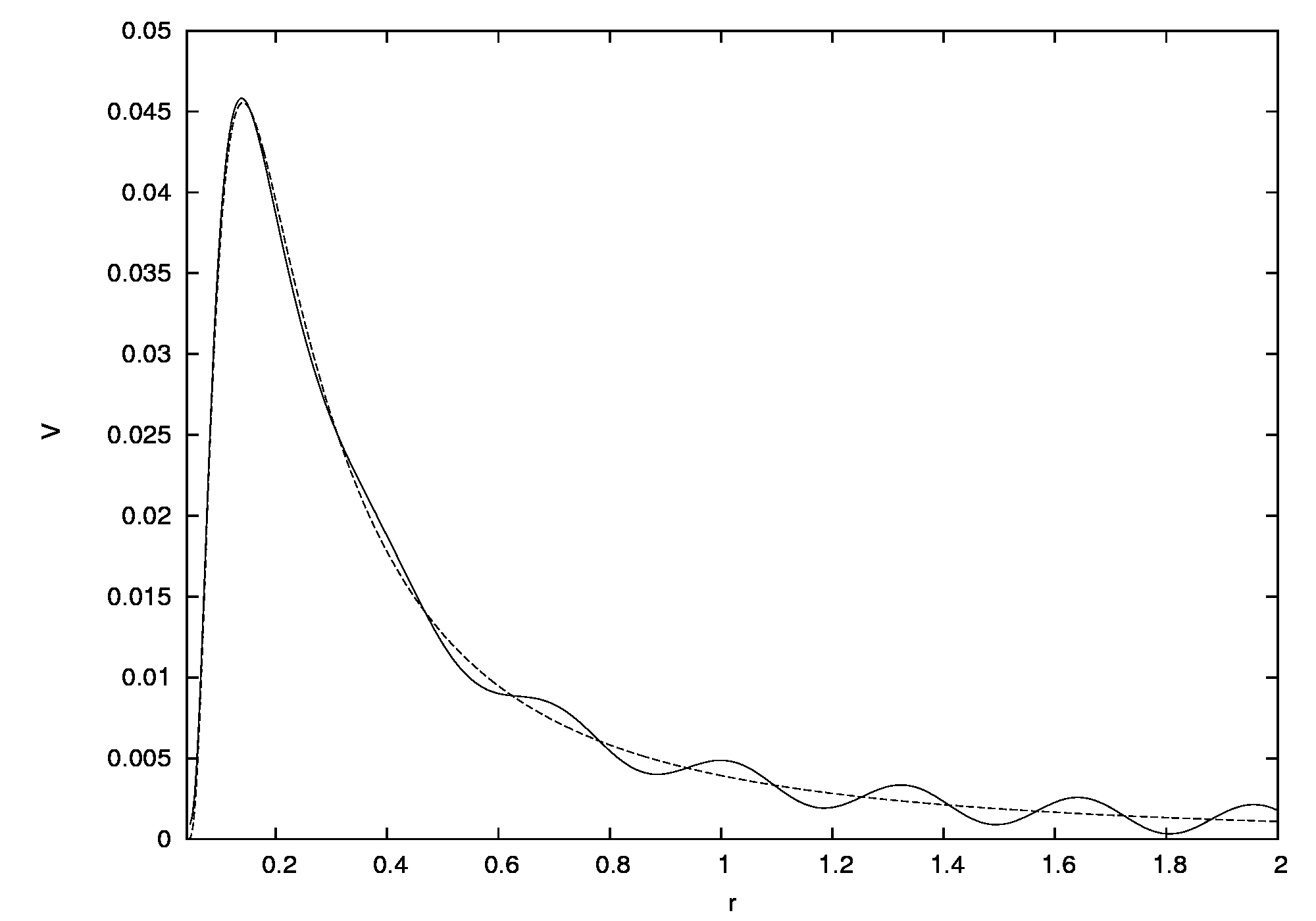}
    \end{minipage}}
\subfloat[Intermediate time]{\label{a25}
    \begin{minipage}[b]{0.48\linewidth}
        \centering \includegraphics[width=0.98\linewidth]{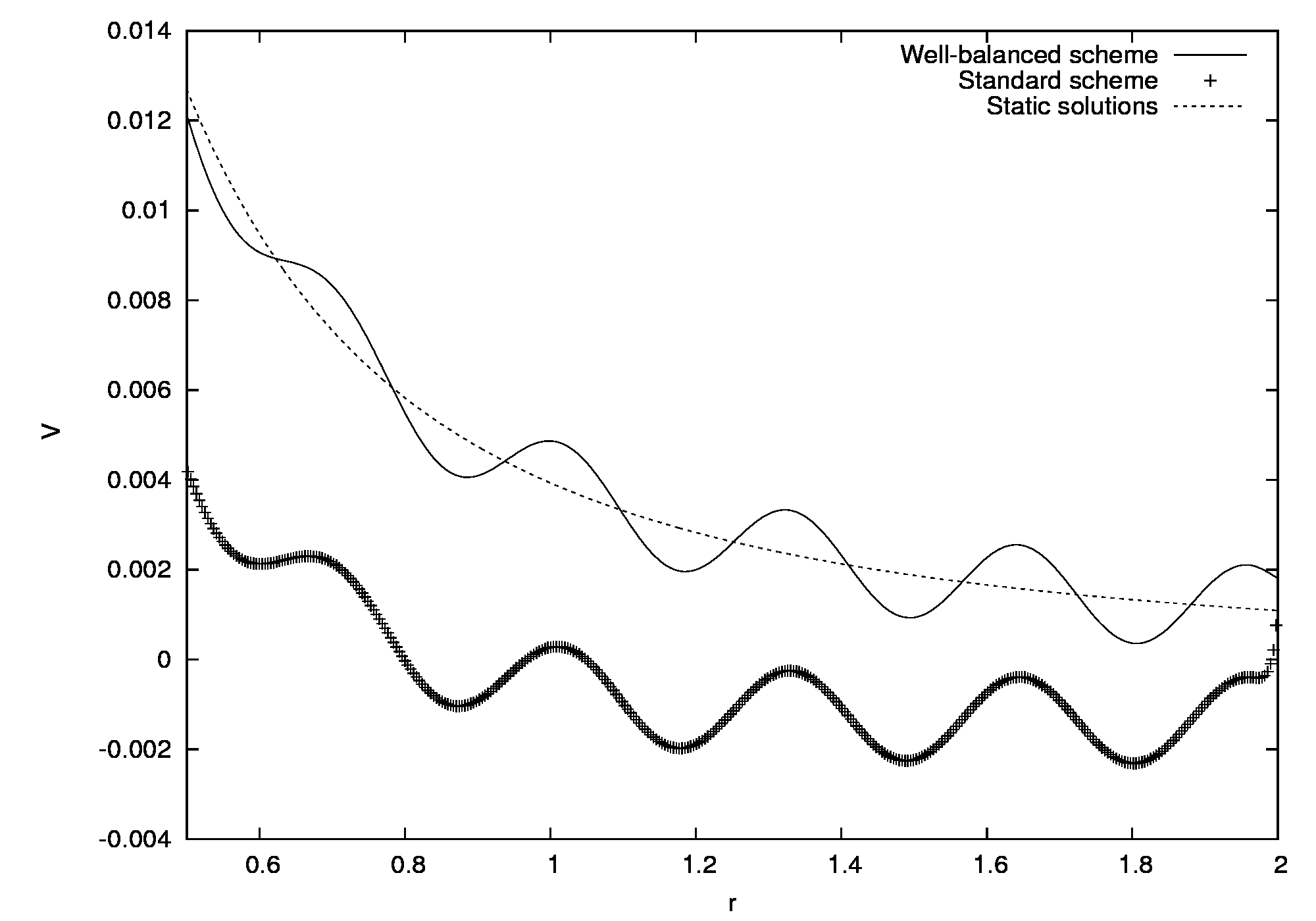}
    \end{minipage}
} \hfill \subfloat[Later time]{\label{a35}
    \begin{minipage}[b]{0.48\linewidth}
        \centering \includegraphics[width=0.98\linewidth]{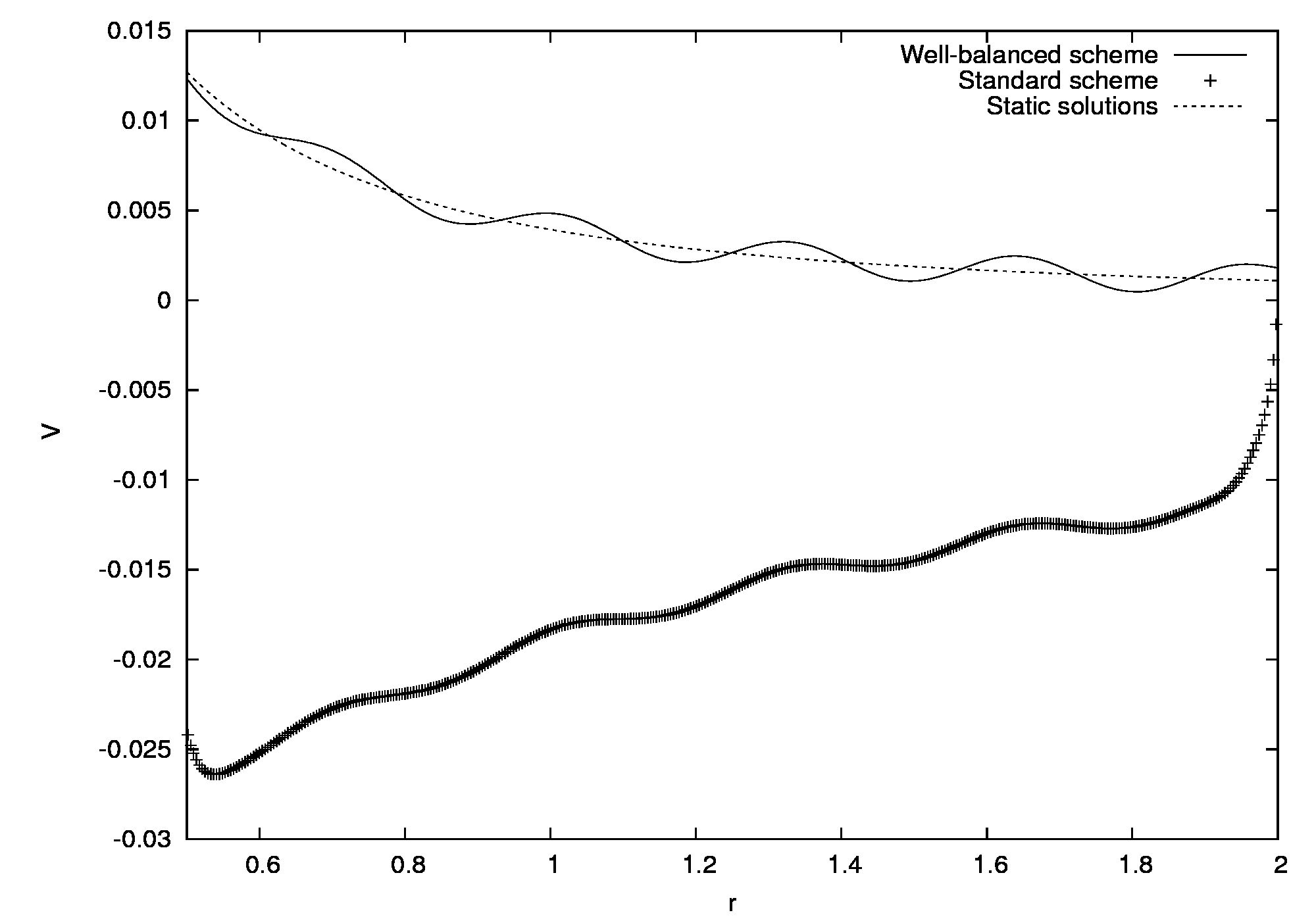}
    \end{minipage}
} \hfill \subfloat[Asymptotic behavior]{\label{a45}
    \begin{minipage}[b]{0.48\linewidth}
        \centering \includegraphics[width=0.98\linewidth]{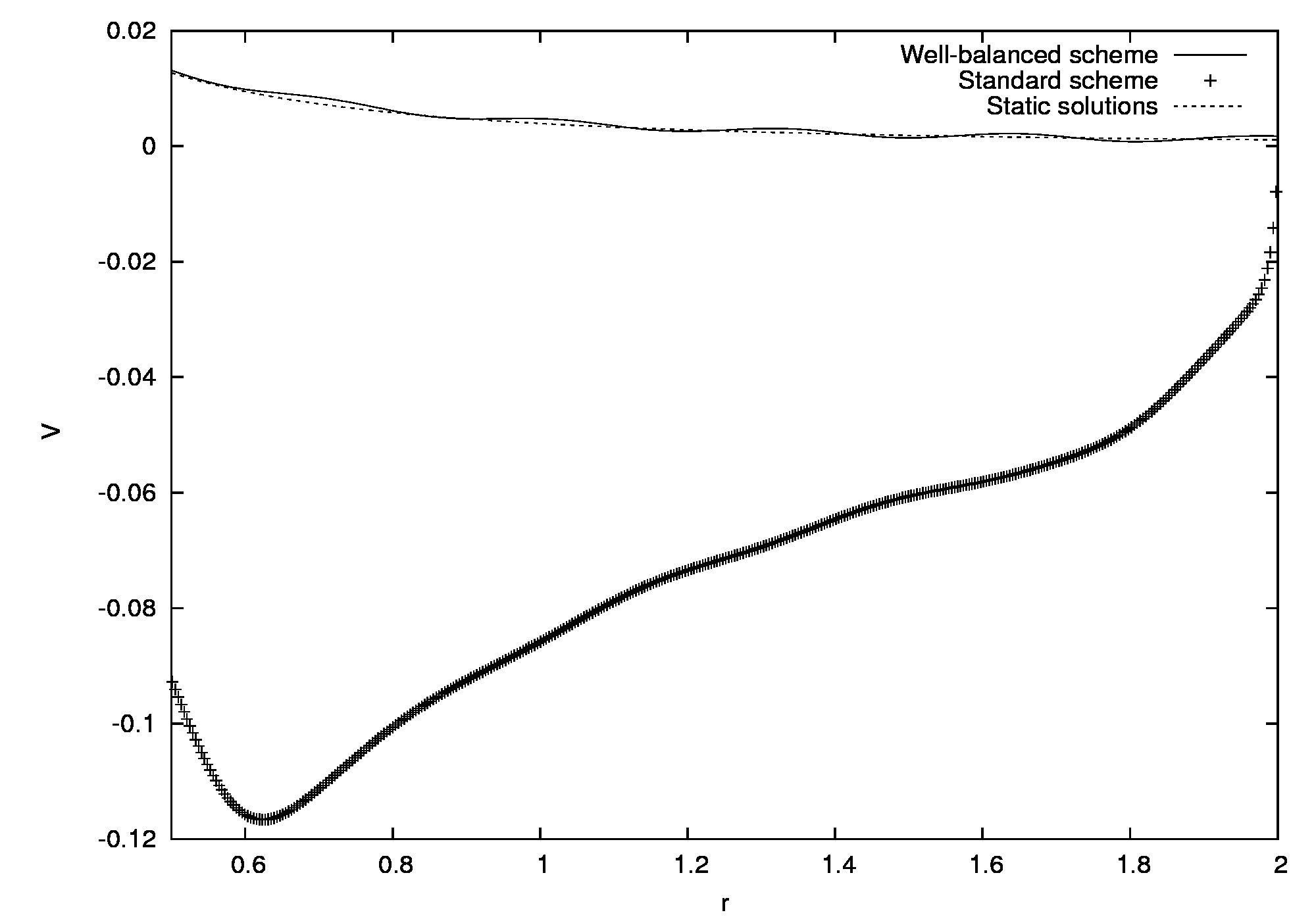}
    \end{minipage}
     } 
\begin{center}
{{\bf Figure 5.1.1.} Numerical solutions with the standard and well--balanced schemes.}
\end{center}
\end{figure}

  \begin{figure}[htp]
  \centering
  \begin {minipage}[ t ]{7.3 cm }
  \centering
  {\includegraphics[height=7cm,angle=90]{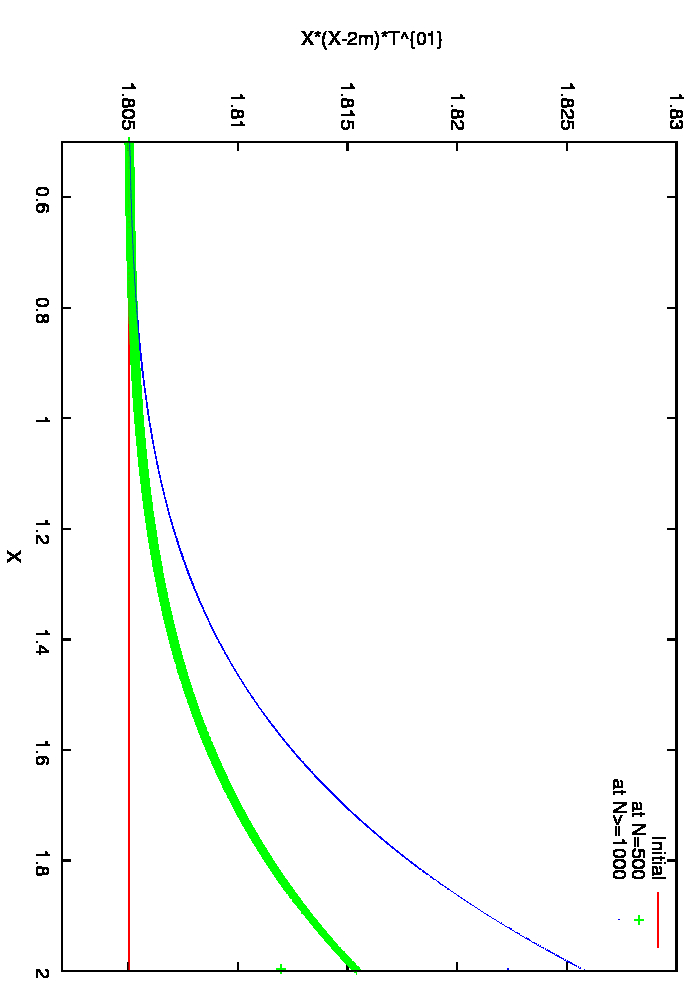}}
\begin{center}
  {(a) $r(r-2m)\,\tT^{01}$ asymptotic values} 
\end{center}
  \end {minipage}
  \begin {minipage}[ t ]{6.3 cm }
  \centering
  {\includegraphics[height=7cm,angle=90]{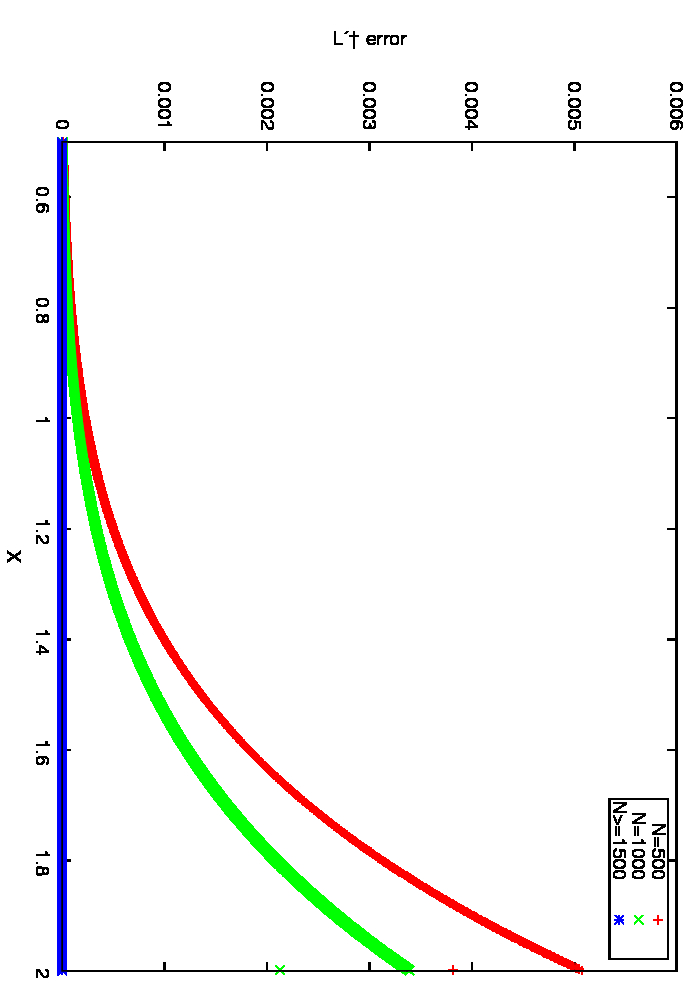}}
\begin{center}
 {(b) $L^1$ error between the numerical and asymptotic solutions} 
\end{center}
 \end {minipage} 
\begin{center}
{{\bf Figure 5.1.2.} Asymptotic solution with the well--balanced scheme.}
\end{center}
 \end{figure}


\subsection{Propagation of a shock/rarefaction pattern} 

We study here initial data containing a shock separating two steady solutions. As we evolve this initial data, the profile of the solution changes and additional waves arise. The initial jump we choose being arbitrary, a full solution to the Riemann problem is generated and both shocks and rarefactions may occur. 
The two steady solutions are defined as follows: the left--hand steady solution has the velocity $V(R) =0.001$ and the density $\rho(R)= 460$,  
while $V(R)=0.004$ and $\rho(R)=480$ for the right--hand solution.  As in the first test, we work in the interval $r_0 = 0.5 < r < R= 2$. 
We use the space mesh size $\Delta r=0.025$ and a CFL number equal to $0.9$. 

Figure 5.2 (a) represents the initial discontinuity separating the two steady solutions, while the other three plots in Figure~5.2 (b), (c), (d) 
show the numerical solution given by the standard and the well-balanced schemes. We observe that the well--balanced scheme produces a sharper solution with a jump of the same magnitude as the initial jump, while the standard scheme has generated a ``spike'' of much larger amplitude. Yet, this unphysical spike is diminishing as time evolves and gets back closer to the amplitude of the well-balanced numerical solution.


\begin{figure}[t!] 
\subfloat[Initial velocity]{\label{a23}
    \begin{minipage}[b]{0.48\linewidth}
        \centering \includegraphics[width=0.98\linewidth]{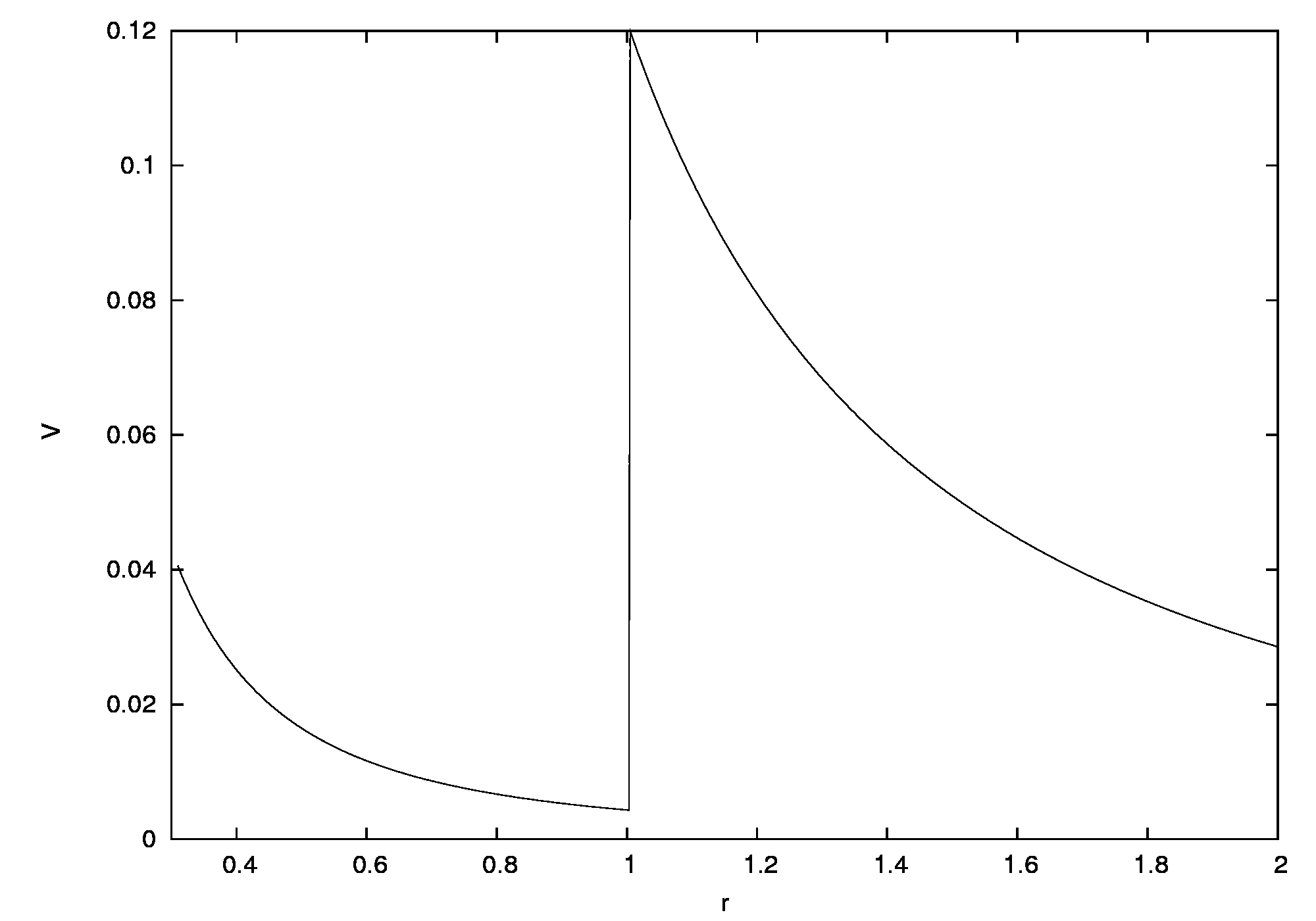}
    \end{minipage}
} 
 \hfill \subfloat[]{\label{a43}
    \begin{minipage}[b]{0.48\linewidth}
        \centering \includegraphics[width=0.98\linewidth]{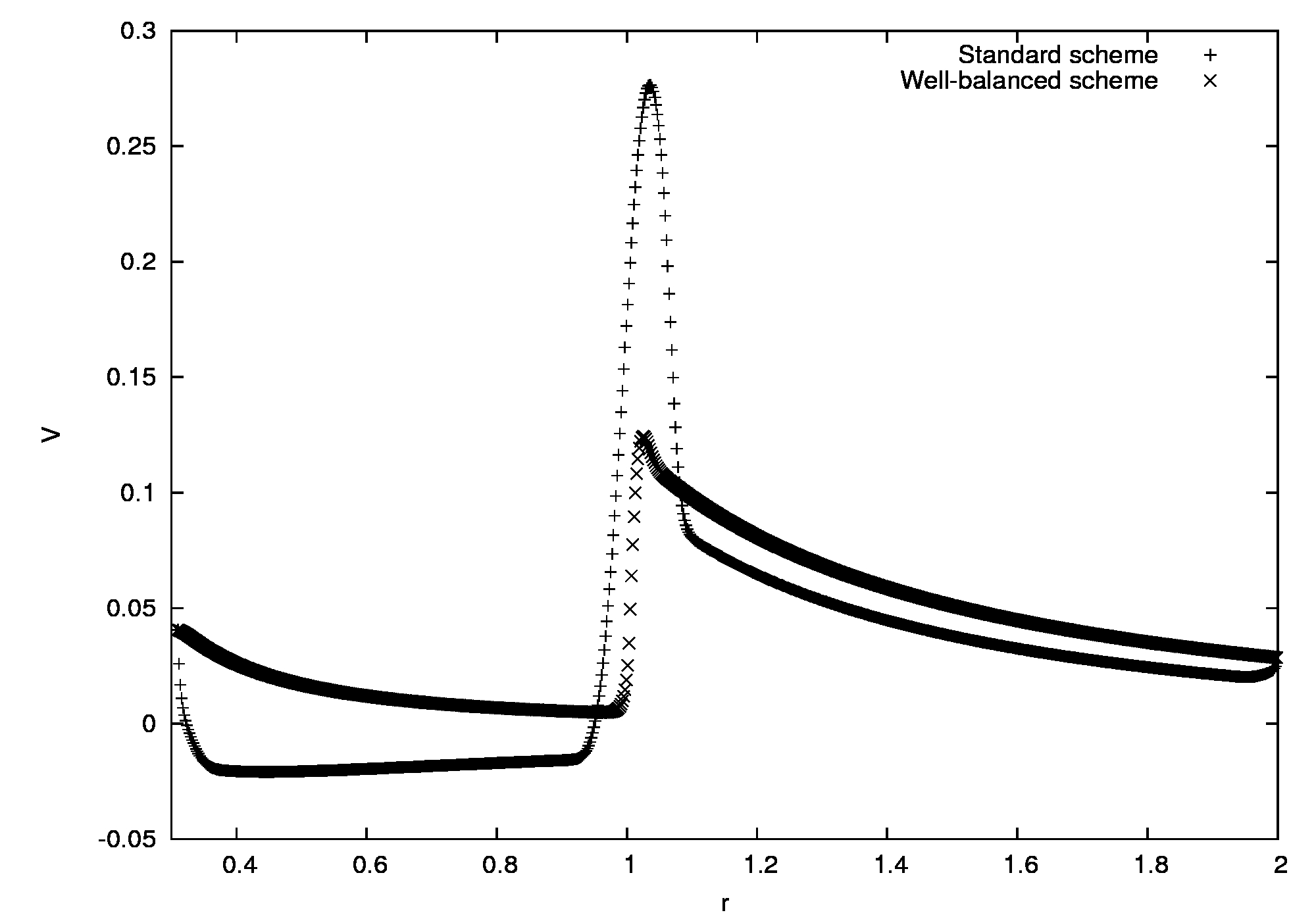}
    \end{minipage}
} \hfill \subfloat[]{\label{a53}
    \begin{minipage}[b]{0.48\linewidth}
        \centering \includegraphics[width=0.98\linewidth]{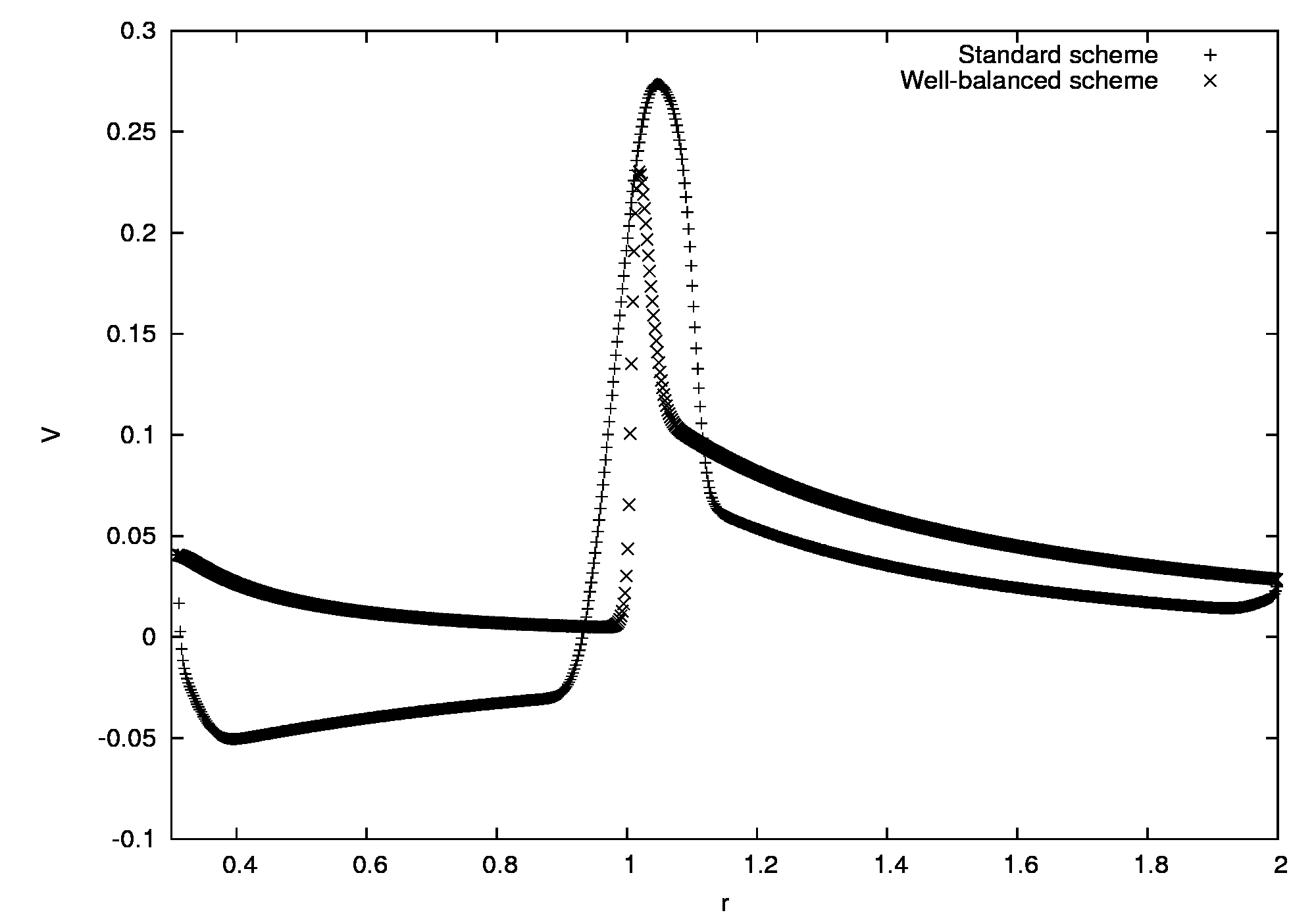}
    \end{minipage}
} \hfill \subfloat[]{\label{a531}
    \begin{minipage}[b]{0.48\linewidth}
        \centering \includegraphics[width=0.98\linewidth]{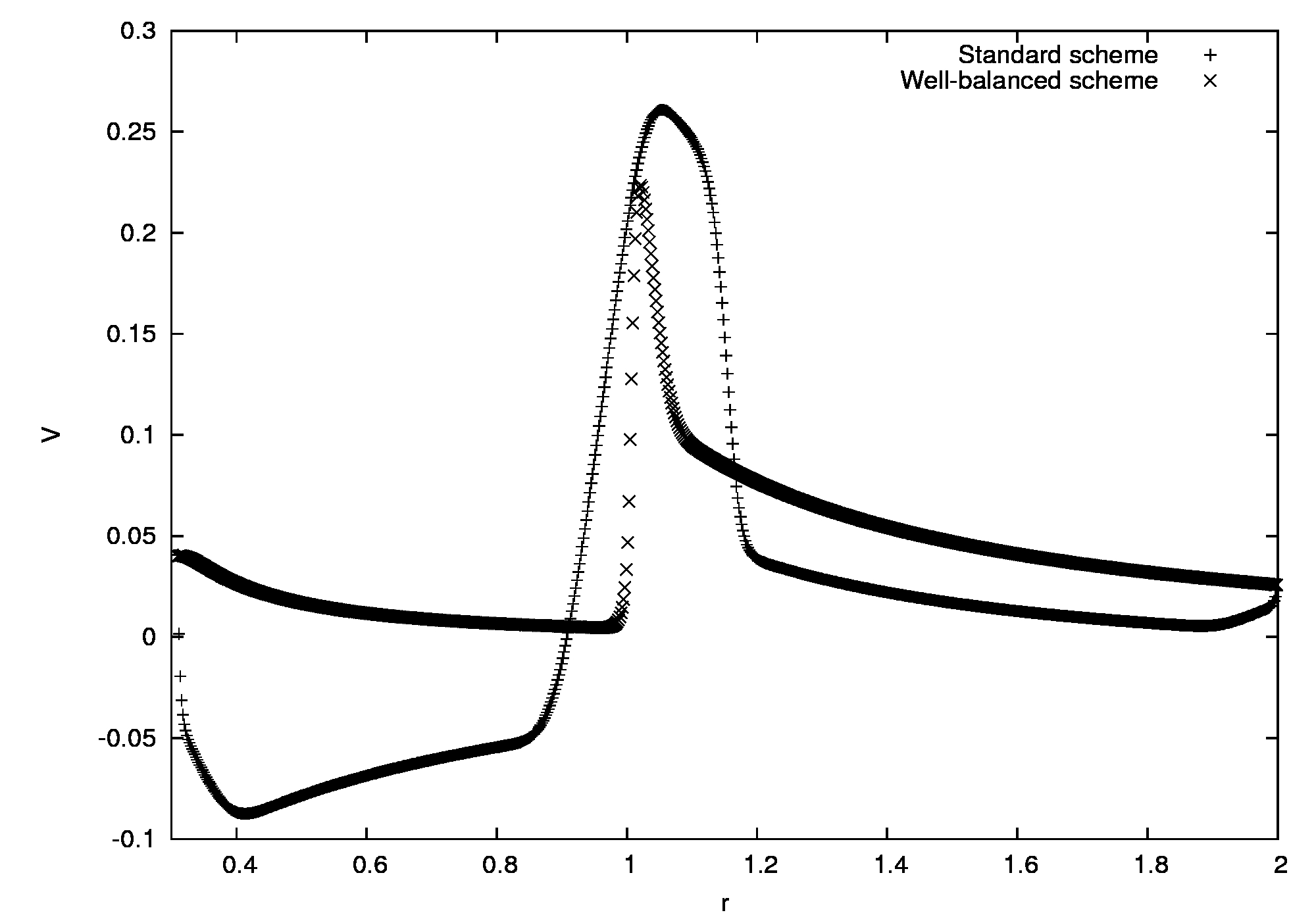}
    \end{minipage}
     } 
\begin{center}
{{\bf Figure 5.2.1.} Propagation of a shock/rarefaction pattern (two schemes).} 
\end{center}
\end{figure}


\subsection{Late--time asymptotic stability of steady solutions}

We have now validated our well--balanced scheme and this motivates us to now apply it in order to 
study the {\sl nonlinear stability} of a given steady solution. The initial data is defined by adding a compactly supported perturbation.  
The velocity and density at the right--hand point of the spatial interval are chosen to be $V(R)=0.001$ and $\rho(R)= 460$, respectively.
We now work in the interval $r_0 = 1.2 < r < R= 2$, and we use the space mesh size $\Delta r=0.035$ and a CFL number equal to $0.9$. 
The time-asymptotic solution correspond to the data $\rho_R=528.0$ and $v_R=0.0011$. 

The numerical results are plotted in Figure~5.3. Observe that the initial discontinuous perturbation evolves, gets smoothed out, and spreads in both directions. In particular, in Figure~5.3~(a) and (b), we compare the numerical solutions for very large times with the initial steady solution, as well as with the steady solution determined from the values $V(R)$ and $\rho(R)$ of the numerical solution. 
Hence, we have demonstrated numerically that solutions converge to steady solutions and, in the present test, the initial perturbation has the effect of modifying the steady solution of reference.  

We observed numerically that the solution reaches another steady state, which we plot on the ssame figure, for the sake of comparison. The late-asymptotic solution is found to be  
  $\rho(R) \simeq 528.0$ and $v(R) \simeq0.0011$.
Furthermore, in Figure~5.4, we have computed the relative numerical error in a log-log scale in terms of the ratio ${1\over \Delta r}$. The convergence rate for the first scheme was found to be $a_1 = - 0.83$, while $a_2=  -1.43$ for the second scheme.


\section{Concluding remarks} 
\label{section:6} 

We have presented a well-balanced scheme for relativistic hydrodynamics posed on a fixed background spacetime and, especially,
 Schwarzschild spacetime. For simplicity, we assumed that the equation of state is given by a linear relationship between the mass-energy density and the pressure. The generalization to other pressure laws is possible; it would lead to ``highly nonlinear'' algebraic expressions, but would not bring any new difficulty for the purpose of this paper.  

To encompass other backgrounds (for instance, Schwarzschild-de Sitter spacetime, with cosmological constant included), the analysis in Section~3 should be revisited. An analogue characterization of steady solutions could be derived without additional conceptual difficulties, so that our method appears to be relevant for a class of background black hole spacetimes.  

Although the proposed finite volume method is currently restricted to problems involving one spatial variable, it 
is of genuine interest for numerical relativity and provides a new tool in order 
to investigate the evolution of {\sl self-interacting matter under symmetry conditions.}  
Indeed, a suitable extension of our method \cite{LMH} 
allows one to deal with the {\sl coupled} Einstein-Euler system, in which the metric itself is an unknown of the problem. This strategy has now been applied to the Einstein equations for spherically symmetric spacetimes 
and should also be useful to investigate $T^2$--symmetric matter spacetimes 
(admitting, by definition, two commuting Killing fields). Specifically, 
we refer to \cite{LMH} for a study of the spherically symmetric, self--gravitational collapse of compressible fluids (investigated earlier by Novak and Ibanez \cite{NI} and Papadopoulos and Font \cite{PapaFont} in a different gauge) and
for the application of the proposed finite volume technique to the formation of trapped surfaces.  

The treatment of the full Einstein system without symmetry assumptions is currently out of the scope of the existing techniques.
An intermediate goal will be to encompass general fluid equations (without symmetry and with general pressure laws) 
and arbitrarily curved four-dimensional background geometries (satisfying Einstein equations).  


\begin{figure}[t!]
\centering \subfloat[Initial data and perturbation (velocity) 
]{\label{a1}
    \begin{minipage}[b]{0.48\linewidth}
        \centering \includegraphics[width=0.98\linewidth]{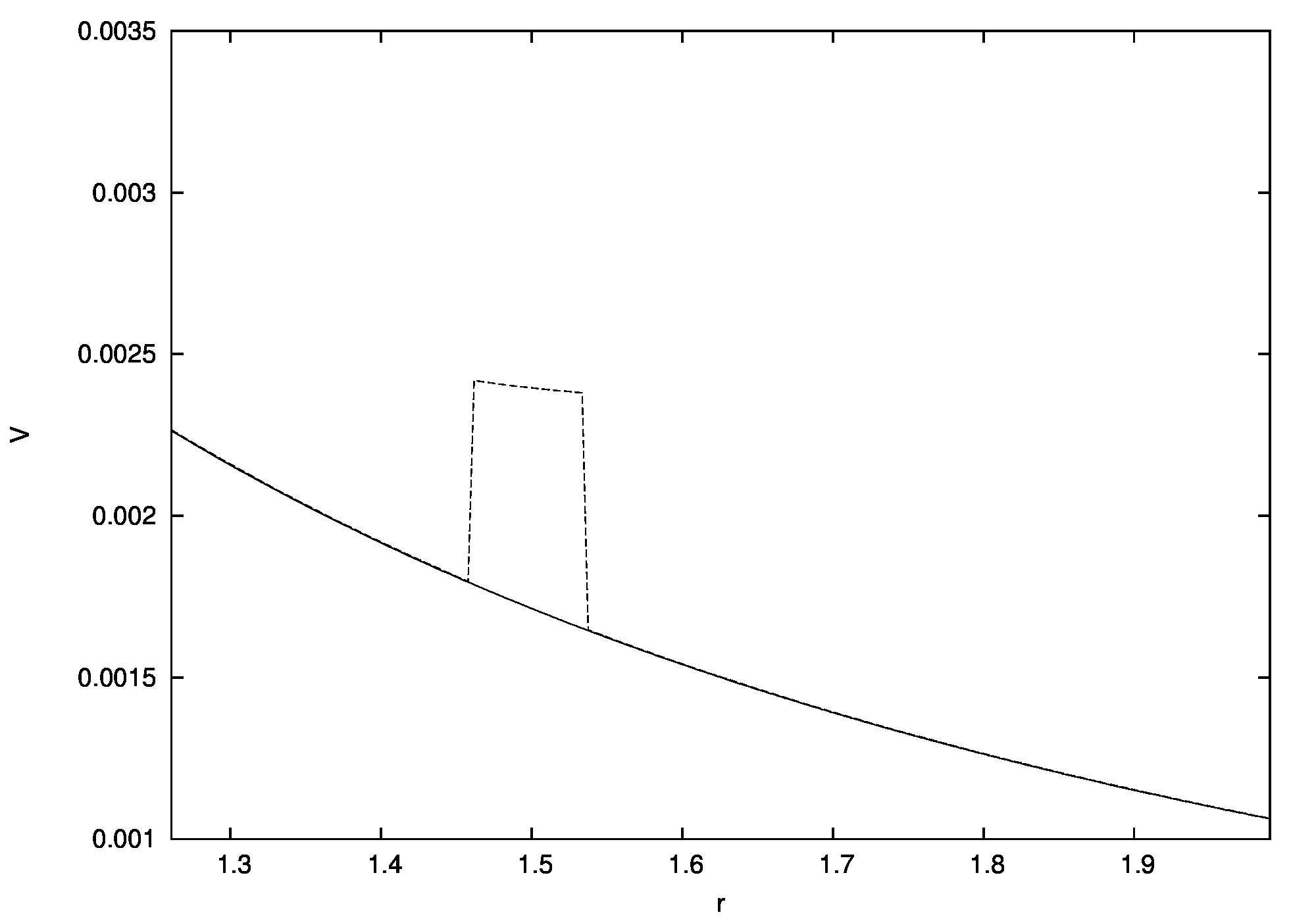}
    \end{minipage}}
\subfloat[t = 0.01]{\label{a2}
    \begin{minipage}[b]{0.48\linewidth}
        \centering \includegraphics[width=0.98\linewidth]{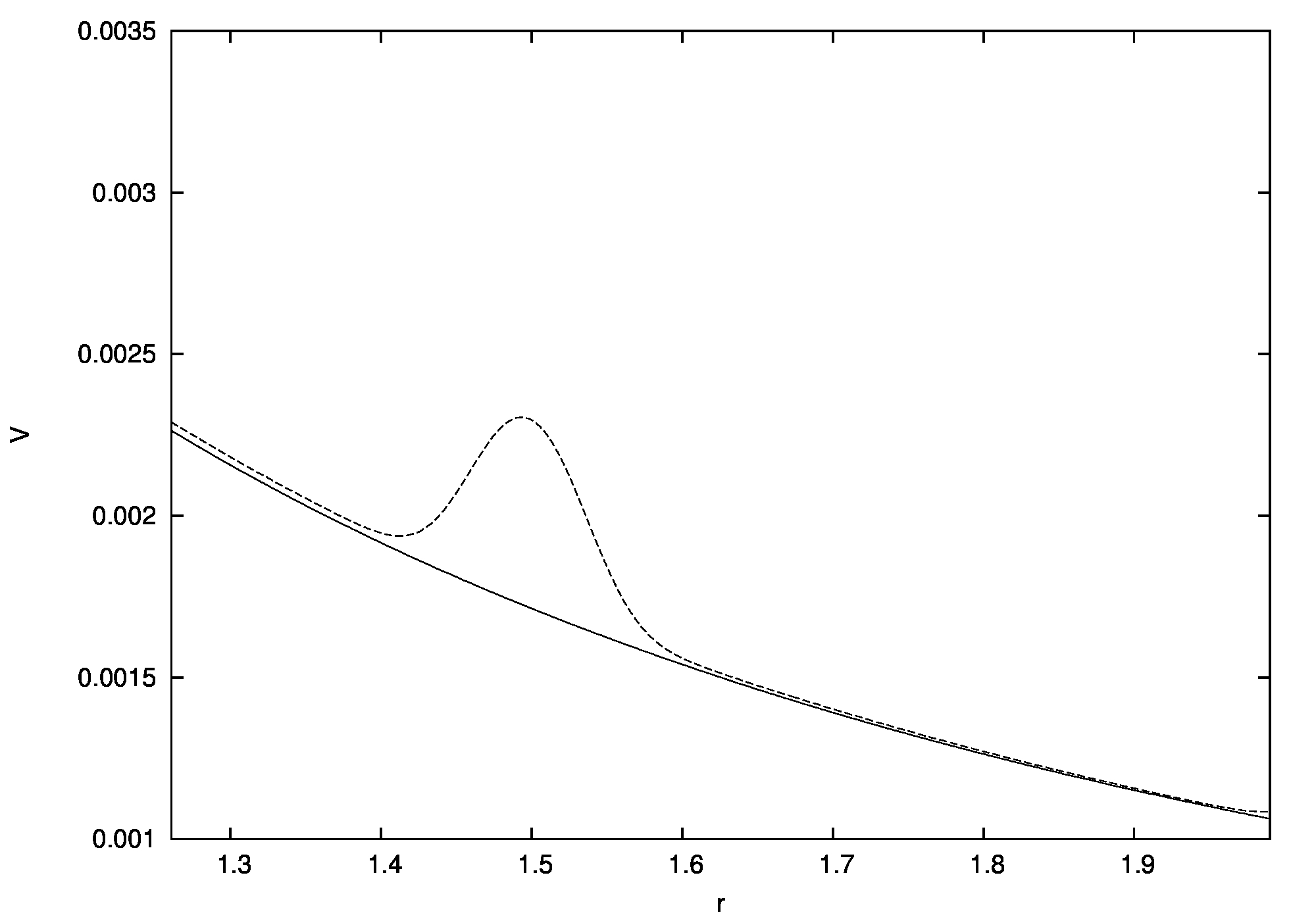}
    \end{minipage}
} \hfill \subfloat[t = 0.05]{\label{a3}
    \begin{minipage}[b]{0.48\linewidth}
        \centering \includegraphics[width=0.98\linewidth]{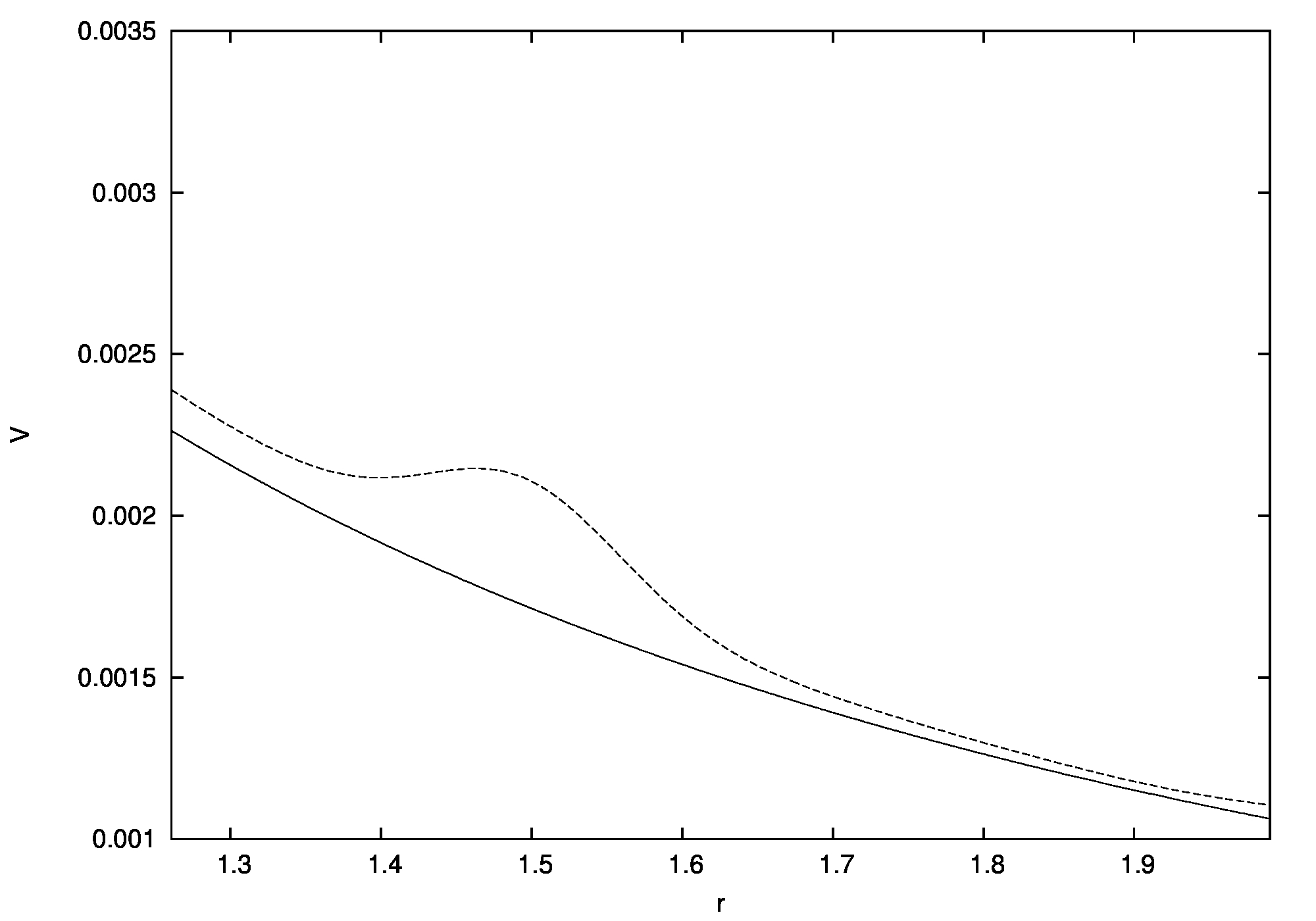}
    \end{minipage}
} \hfill \subfloat[t = 0.08]{\label{a4}
    \begin{minipage}[b]{0.48\linewidth}
        \centering \includegraphics[width=0.98\linewidth]{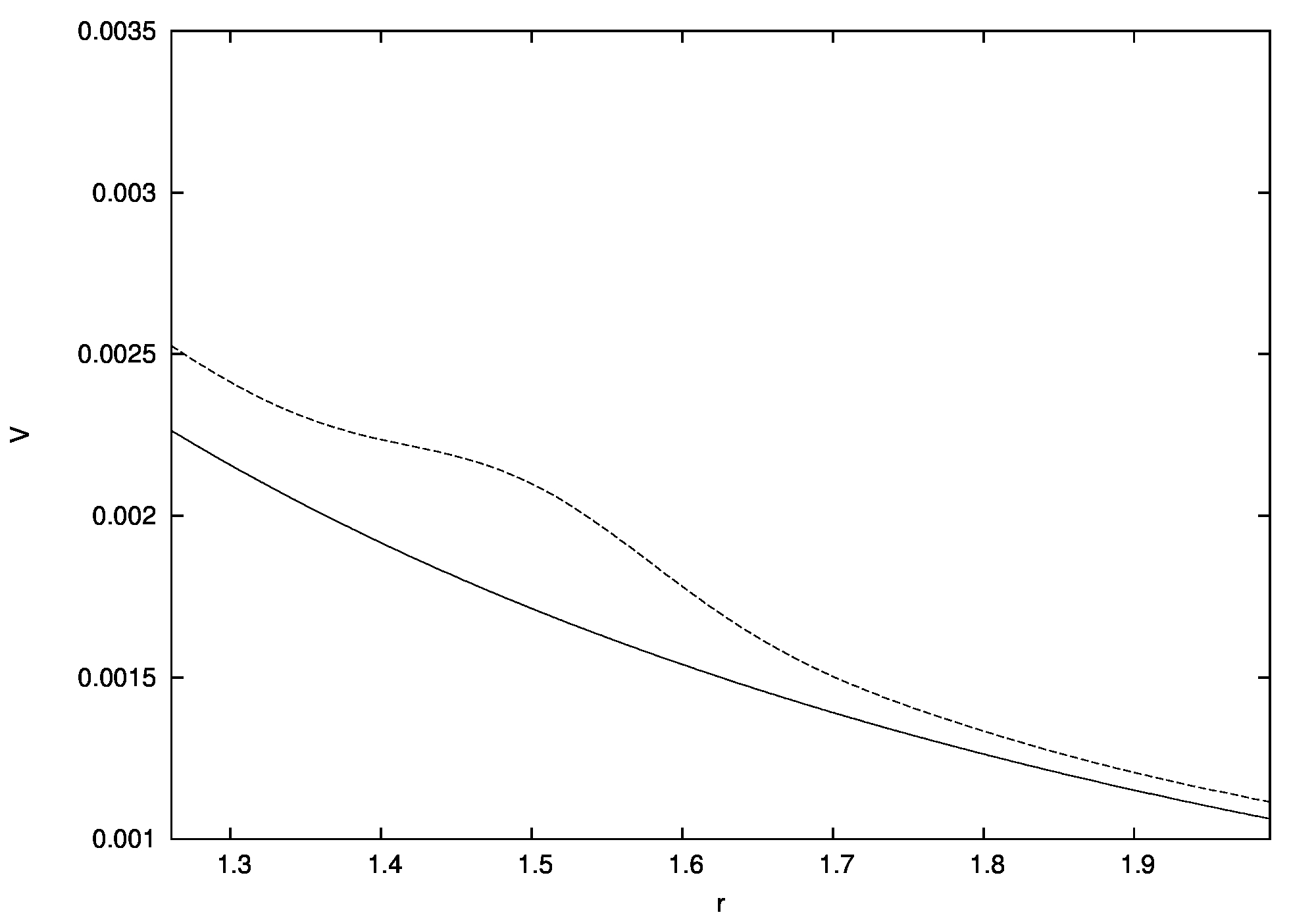}
    \end{minipage}
     } \hfill \subfloat[t = 0.5]{\label{a5}
    \begin{minipage}[b]{0.48\linewidth}
        \centering \includegraphics[width=0.98\linewidth]{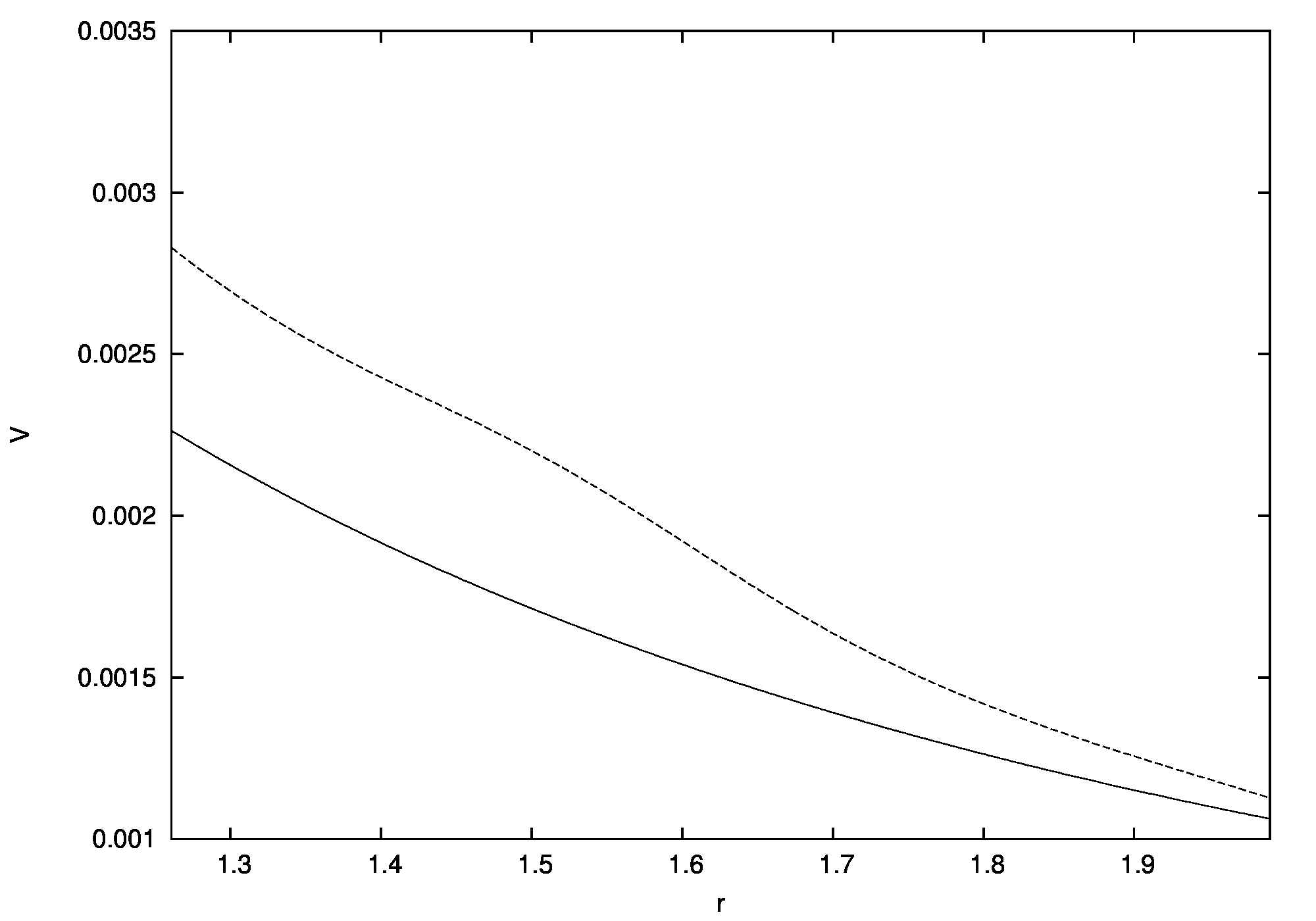}
    \end{minipage}
} \hfill \subfloat[t = 1.55]{\label{a6}
    \begin{minipage}[b]{0.48\linewidth}
        \centering \includegraphics[width=0.98\linewidth]{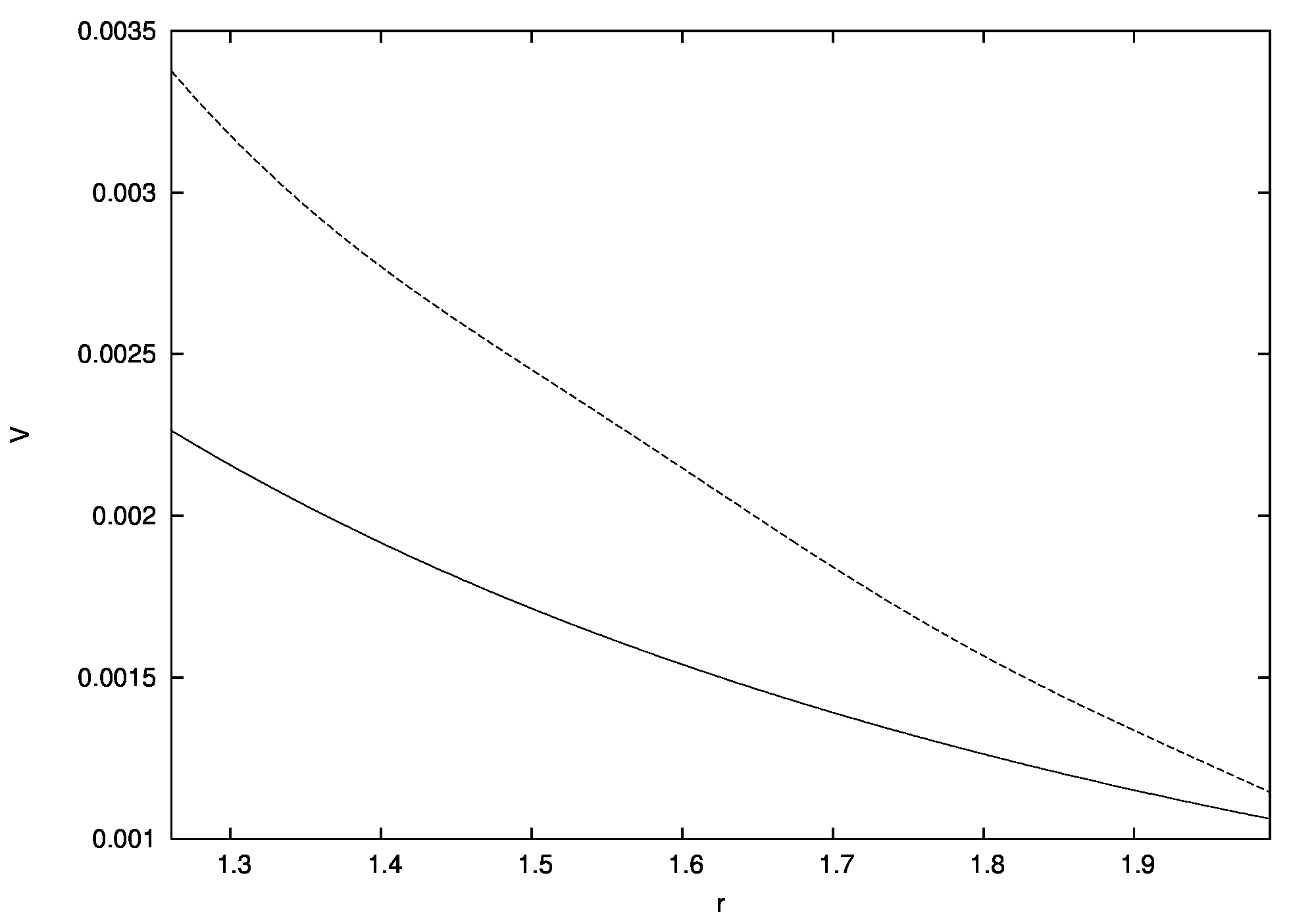}
    \end{minipage}
     }

\begin{center}
{{\bf Figure 5.3.} Late--time behavior of a perturbed steady solution.}
\end{center}
\end{figure}

\begin{figure}[t!]
\centering {\label{Error}
    \begin{minipage}[b]{0.70\linewidth}
        \centering \includegraphics[width=0.98\linewidth]{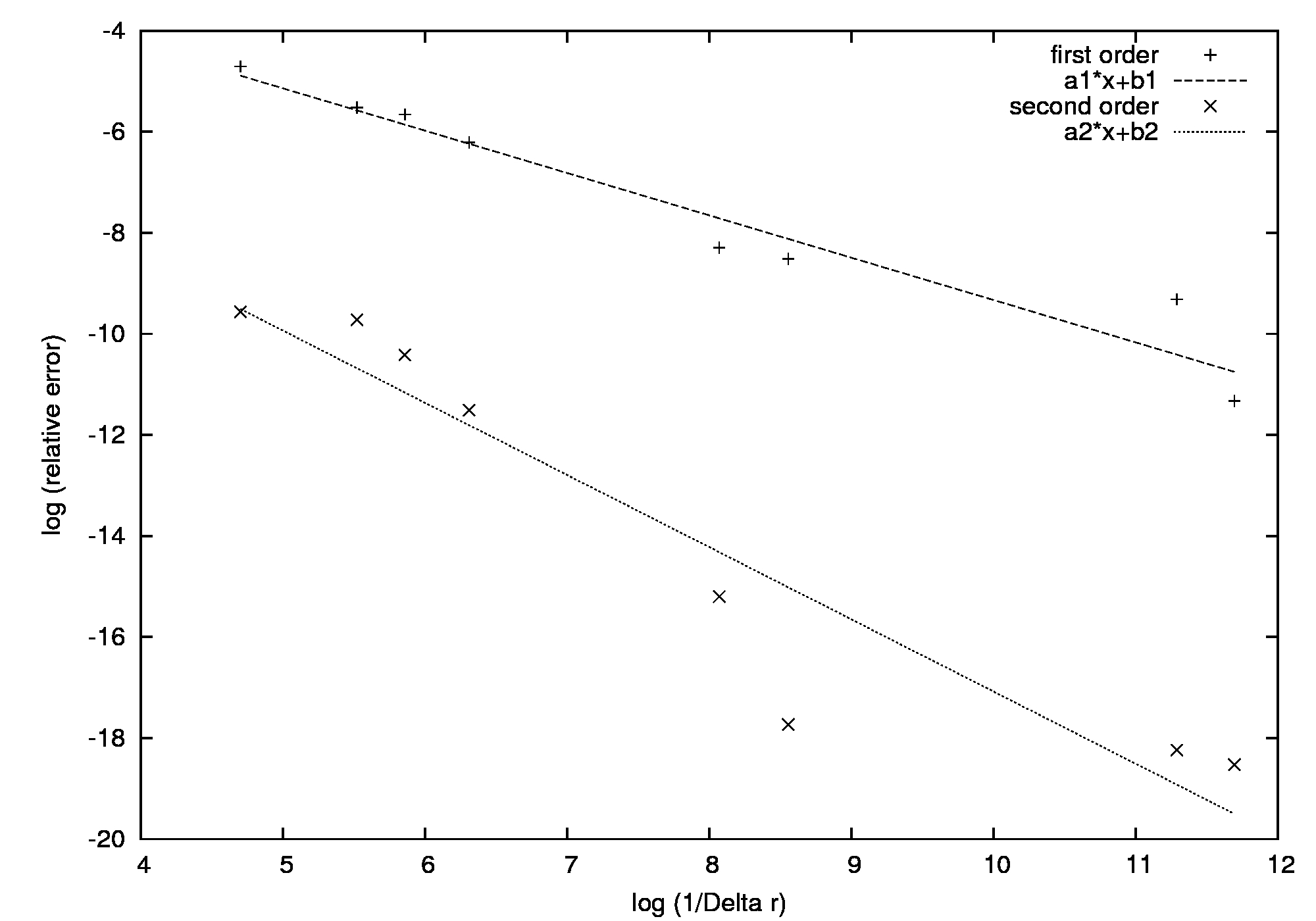}
    \end{minipage}}
\begin{center}
{{\bf Figure 5.4.} Data points and fitted regression lines for the first and second order schemes
\newline 
with $a_1 = -0.83$, $a_2= -1.43$.}
\end{center}
\end{figure}


\section*{Acknowledgments}
 
The authors are very grateful to J.A. Font, J. Novak, and G. Russo for their comments on a preliminary version of this paper.  
The first author was supported by the Centre National de la Recherche Scientifique. 
The authors were  supported by the Agence Nationale de la Recherche through the grants ANR 2006-2--134423  
and ANR SIMI-1-003-01.


\end{document}